\documentclass[fleqn,usenatbib]{mnras}

\usepackage{txfonts}

\usepackage[T1]{fontenc}
\usepackage{ae,aecompl}

\usepackage{graphicx}    
\usepackage{amssymb}    
\usepackage{array}
\usepackage[dvipsnames]{xcolor}
\usepackage{booktabs}
\usepackage{longtable}
\usepackage{supertabular,booktabs}
\usepackage{threeparttable}
\usepackage[bottom]{footmisc}
\usepackage[caption=false]{subfig}

\newcommand{\Msun}{\ensuremath{M_{\odot}}}

\newcommand{\M}{\ensuremath{M_{\odot}}}

\newcommand{\somaked}[1]{\textcolor{black}{#1}}

\title[Evolution of galaxies in the Coma Supercluster]{Evolution of galaxies in groups in the Coma Supercluster}

\author[Seth and Raychaudhury]{Ruchika Seth$^{1}$\thanks{E-mail: ruchika@iucaa.in}
and
Somak Raychaudhury$^{1,2,3}$\\
$^{1}$Inter University Centre for Astronomy and Astrophysics, Pune 411007, India\\
$^{2}$Department of Physics, Presidency University, 86/1 College Street, Kolkata 700073, India\\
$^{3}$School of Physics and Astronomy, University of Birmingham, Birmingham B15~2TT, UK\\
}

\date{Accepted XXX. Received YYY; in original form ZZZ}

\pubyear{2019}

\begin{document}
\label{firstpage}
\pagerange{\pageref{firstpage}--\pageref{lastpage}}
\maketitle

\begin{abstract}
We take a close look at the galaxies in the Coma Supercluster and assess the role of the environment (in the form of cluster, group and supercluster filament) in their evolution, in particular examining the role of groups. We characterise the groups according to intrinsic properties such as richness and halo mass, as well as their position in the supercluster and proximity to the two rich clusters, Abell 1656 (Coma) and Abell 1367. We devise a new way of characterising the local environment using a kernel density estimator. We find that apart from the dominant effects of the galaxy mass, the effect of the environment on galaxies is a complex combination of the overdensities on various scales, which is characterised in terms of membership of groups, and also of the position of the galaxy on filaments and their proximity to the infall regions of clusters. Whether the gas can be turned into stars depends upon the level of pre-processing, which plays a role in how star formation is enhanced in a given environment. Our results are consistent with gas accreted in the cold mode from the filaments, being made available to enhance star formation. Finally, we show that the Abell~1367 end of the supercluster is in the process of assembly at present, leading to heightened star formation activity, in contrast with the Coma-end of the filament system.
\end{abstract}



\section{Introduction}
\label{section:Section-1}

Galaxies can evolve passively without much interaction with their surroundings, gradually consuming their own gaseous content, the rate of star formation diminishing as the fuel is consumed. On the other hand, the evolution of galaxies has been shown to respond to their local environment, whether they reside in groups or clusters, or navigate the cosmic web. This could be related to the supply of gas in various thermal states, the removal of the gas due to direct (stripping) or indirect (tidal) processes, or the occurrence of encounters that could inspire star formation or AGN activity. Nature and nurture are thus both important in the evolution and transformation of galaxies.


One popular scenario is that most of the giant galaxies irrespective of their environments are found to evolve passively, thus indicating that the mass of the galaxy  is the most dominant factor rather than environment. For example, \citet{Blanton2006} finds that where star formation properties of massive galaxies relate to over-densities over scales of a few Mpc, but do not show correlation with overdensities on larger scales, concluding that galaxy evolution is related to the mass of the host galaxy-scale or group-scale dark-matter haloes than the larger-scale environment.

On the other hand, early research involving samples of galaxies spanning a large range of mass and luminosity, such as that of \citet[e.g.][]{Dressler1980}, had found that the dominant morphology of galaxies evolves with increasing local density of galaxies, e.g. as in a radial trend moving from the outskirts of a galaxy cluster to its core. This continues to be found in various forms in the ever-growing large-volume samples of galaxies, \somaked{thus indicating that the large-scale and small scale environments of a galaxy are important factors to consider, along with its mass, in its evolutionary history.}

The filaments of the cosmic web are believed to have formed hierarchically from smaller primordial structures, connecting and feeding the proto-galaxies through filamentary streams \citep{Keres2009}, which in turn had formed due to the collapse of dark-matter at a relatively earlier epoch. These filaments, in particular those connecting clusters, can form networks of large megaparsec-scale structures of dark-matter and galaxies \citep[e.g.][]{porter2005}, often forming a ordered and coherent velocity field along them, defining the flow of galaxies \citep[e.g.][]{aragon2016}.

\somaked{Thus, these filaments \citep{Lietzen2012,Kuutma2017,Kleiner2017}
play an important role in galaxy transformations.
As galaxies approach the infall regions of clusters at the crossroads of the cosmic web, there is indication of enhancement of star formation in galaxies in clusters fed by filaments \citep[e.g.][]{porter2007,porter2008,Mahajan2012}.
The filaments also enable gas flow onto galaxies and groups of galaxies embedded in the cosmic web. The gas can flow from the filaments in cold mode, which can aid star formation under appropriate conditions, or in hot mode, where the gas is at an elevated temperature and entropy, which is much harder to convert into stars, even though the conditions might be otherwise conducive
\citep[e.g.][]{Balogh1999,Keres2009,Noguchi2018}.}

As an example, from a systematic survey of the infall region of a sample of clusters, \citet{Haines2018} found that about 35-50\% mass accretion on massive clusters takes place between $z\!\sim\!0.2$ and the current epoch. This tidally affects the galaxy population on the outskirts of clusters, which has various consequences on the evolution of galaxies.

Apart from the location of galaxies on the large-scale web of filaments, the environment of galaxies in systems such as groups is expected to play a role in their evolution as well. At the present epoch, most of the stars are in poor group environments, with less than 5\% being in rich clusters \citep[e.g.][]{Eke2005}, making groups worthy of exploration in order to study galaxy evolution. In galaxy groups, several studies \citet[e.g.][]{Wetzel2012,Mahajan2013,Cybulski2014} find that `pre-processing' can lead to faster quenching of galaxies, compared to those in the field. 

The term ``pre-processing" refers to the enhanced transformation of galaxies  prior to their assimilation into the clusters, in their journey on the cosmic web. Certain environments (e.g. groups or sub-clusters) are sufficiently prone to such interactions, being relatively underdense. Galaxies in groups tend to have lower relative velocities, which enhances dynamical friction, bringing them closer for enhanced interaction times, which assists morphological transformations \citep[e.g.][]{Haines2007}. The degree of pre-processing seems to be correlated with the group properties such as group halo mass and virial radius \citep{Wetzel2012,Woo2013,Weinmann2006}. These seem to indicate that groups are not merely scaled-down clusters. 

In the above scenario, while the individual galaxies are vulnerable to stripping processes and are seldom able to hold on to their gaseous content, owing to their shallow potential wells, the situation is entirely different for galaxies in groups. Groups have a deeper dark-matter potential compared to that in galaxies, and have a stronger hold on the gas. On the other hand, these member galaxies are vulnerable within their own group environments, which offer increased dynamical friction, leading to enhanced galaxy-galaxy interactions, aiding processes which contribute to `pre-processing' (see \S\ref{subsection:preprocess}).

If the predominant environmental factor driving galaxy transformation is the overdensity of neighbours, would it matter if galaxy belongs to a group or not when it is falling into the cluster?
Thus it is important to consider the properties of the groups or clusters to which a galaxy belongs, along with its position in the cosmic web and the supercluster filaments through which it may have passed or in which it might be currently embedded.

Various methods  have been adopted to characterise the environment of galaxies 
For example,
\cite{Muldrew2012} have collected 20 such environmental parameters and classified them into two classes: ones resulting from nearest-neighbour algorithms, which trace the underlying density of high mass dark-matter haloes, and those from fixed-aperture algorithms, which are better probes of the super-halo scales. 

In this paper, we explore one such part of the cosmic web which offers a wide range of environments: the Coma Supercluster, a very well-studied system in our local universe. Located at a distance of about 100 Mpc from us, it consists of two rich clusters, namely the Coma Cluster (Abell~1656) and Abell~1367, connected via a rich network of filaments, in which are embedded a large number of groups and galaxies that can be closely studied due to the proximity of the system  \citep[among others,][]{Williams1981AJ, Chincarini1983A, Fontanelli1984, Gavazzi1986, Gavazzi1989,  Rines2001, Boss1995, Mahajan2010, Gavazzi2011, Mahajan2018}. 

Although the two Abell clusters are assigned the same richness ($R=2$) in the Abell catalogue, a detailed look reveals them to be very different. The Coma cluster is dynamically relaxed, predominantly consisting of early-type galaxies, while Abell~1367 appears to be dynamically young with a higher content of late-types \citep{Doi1995}.

Several studies point to evidence of pre-processing on the filaments before being assimmiliated into the clusters in the case of Coma supercluster \citep{Cybulski2014,Mahajan2010,Gavazzi2010,Mahajan2018}. The infall regions of both clusters show the dwarf galaxies to be evolving from blue cloud to red sequence while assembling through filaments onto the cluster, present in groups and out of them \citep{Mahajan2011}.

One of the key issues we would like to examine
here is the link between the properties of these two clusters and their immediate surroundings. We would like to know how the environment nurtures the galaxies, as individuals or as group members, embedded in their environments. We would also like to investigate how the filament environment alters the properties of groups embedded in them, as they are being accreted onto clusters, and what role the dark-matter haloes of the groups play in this process. Do the groups found elsewhere on the cosmic filament differ from the ones at the vicinity of the cluster?

This paper is arranged as follows: we describe our data in the next section, and characterise the various categories of environment for our sample galaxies in \S{3}. In \S4 we present various properties of galaxies, namely their  specific star formation, stellar mass, the AGN fraction (optical and radio) and morphology, as a function of environment, paying particular attention to group membership. 
We discuss these results in  \S{5} in terms of models of evolution, and summarise the conclusions in \S{6}.

Throughout the paper we adopt the standard $\Lambda$CDM Cosmology, assuming a flat universe with $\Omega_\Lambda\! =\! 0.7$, $\Omega_M\! =\! 0.3$ and $H_0 = 73 $ km~s$^{-1}$ Mpc$^{-1}$.

\section{Observational Sample}
\label{section:Section-2}

The following section describes the various catalogues used for the population analysis of our sample.

\subsection{SDSS spectroscopic sample}
The Sloan Digital Sky Survey (SDSS) is a flux-limited spectroscopic and photometric galaxy survey, carried out by a 2.5 m telescope at Apache point observatory. 
All galaxies are taken from the spectroscopic sample of the SDSS Data Release 13 (DR13; \citet{Albareti2017}). This sample contains 6958 galaxies spanning the entire Coma supercluster within the range 
$165^\circ \leqslant$ RA $\leqslant 210^\circ$, 
$14^\circ \leqslant$ Dec. $\leqslant 34^\circ$, 
4500 km~s$^{-1}$ $\leqslant$ cz $\leqslant$ 9500 km s$^{-1}$. The mean redshift of the two Abell clusters on either side of the central filaments in this supercluster is the Coma cluster (Abell~1656)  at 6950 Km~s$^{-1}$ and Abell~1367 at 6450 Km~s$^{-1}$. Here we have covered a wider projected area compared to some previous surveys \citep[e.g.][]{Mahajan2010}, in order to sample all filaments connected to both clusters.  The spectroscopic limit of our sample goes down to an apparent magnitude of 17.77. This gives us a total of 6958 galaxies. We further employ an absolute magnitude cut at $M_r\!=\!-17.25$, which corresponds to an apparent magnitude of ~17.77 at the mean redshift of the Coma supercluster. A total of 6609 galaxies remain in our final sample. This will be referred to our 'base sample' for future reference.

\subsection{Derived Spectroscopic Parameters}
The physical properties of the galaxies e.g. their Star Formation Rate (SFR) (\M{}yr$^{-1}$) and Specific Star Formation Rate (sSFR) (yr$^{-1}$) were obtained from \cite{Brinchmann2004}
In their work, the values of SFR and sSFR were obtained by employing a Bayesian model, to the spectra obtained from the central $3^{\prime\prime}$ fibre, which involves the emission and absorption lines in the spectra, as well as the continuum, obtained for each of the galaxies. The fitting returns a likelihood distribution for SFR and sSFR. We take the median value of this distribution as our sSFR estimate.

We classify a galaxy as star-forming if the specific star-formation rate, log(sSFR) > -11. The catalogue also indicates were a galaxy might harbour an AGN, using the optical BPT diagnostic \citep{BPT} and further classifies the optical AGN, if present into its sub-types, e.g., AGN, broadline or a composite (AGN + broadline) for each of the galaxies listed in the catalogue. On performing a cross-match we get sSFR for 6432 galaxies out of a total of 6609 in the `base-sample', which includes both galaxies belonging to groups (`Group Galaxies', membership defined below) and not (`Non-Group Galaxies'). We retain the non-matched galaxies in the sample by flagging them as non-star forming.

\subsection{Group Catalogue}
\label{subsection:group}
The \citet{Yang2007} group catalogue is an optically-selected, volume-limited group catalogue which is based on the SDSS DR7 galaxy catalogue (which contains all the galaxies that are used in the region). The catalogue is based on an iterative halo-based group finding algorithm, using an adaptive filter model which considers the  properties of the dark-matter haloes and their distribution in  phase space. \citet{Yang2007} assigns a group membership to every galaxy. Each galaxy in SDSS thus belongs to a dark matter halo. Our group sample is limited to those with four or more member galaxies. Table~\ref{tab:Tab-1} lists the parameters for the two clusters Yang et al group sample while Table~\ref{tab:Tab-2} shows a sample of all other Yang et al groups present in our chosen Coma Supercluster region. The `Group Galaxies' sample has 2242 galaxies, in 109 groups which includes the two clusters. In our sample, we have 4367 `non-group' galaxies.

\subsection{Radio AGN}
\label{subsection:radio AGN}
The \citet{Best2012} catalogue provides a total of 18,286 galaxies detected at 1.4 GHz radio, by combining the SDSS~DR7 with the NVSS and FIRST catalogues. The catalogue classifies the source of radio emission as being from star formation, or in the form of radio-quiet or radio-loud AGN. Here we use the ``radio-loud" subsample. \somaked{In order to separate the radio-loud population from the ones primarily dominated by star-formation, \citet{Best2012} used three independent diagnostic tools, namely the BPT \citep{BPT} criterion (where line-ratios of certain diagnostic optical emission lines), 
the position of a galaxy in the D4000 \& L$_{rad}$/M \citep{Best2005} and L$_{H_{\alpha}}$ \& L$_{rad}$ planes (both of which are based on the premise that the radio-active AGN would be segregated from star-forming population due to high radio luminosity offered by the AGN).
Results from all three diagnostics were further weighted and assessed to compile the radio AGN sample.}

The radio-loud subsample is further classified by  \citet{Best2012} into high excitation (HERG) and low excitation (LERG) radio galaxies, which are indicative of the rate of the radiative accretion. While LERG (quasar mode) are mostly associated with groups and high mass hot haloes, often found in clusters, leading to inefficient (< 1\%) accretion rates, the HERG (radio-mode) galaxies are associated with the accretion of cold gas and are mostly hosted in relatively small galaxies with mergers and relatively younger stellar populations.

\begin{figure*}

\centerline{\includegraphics[width=\paperwidth]{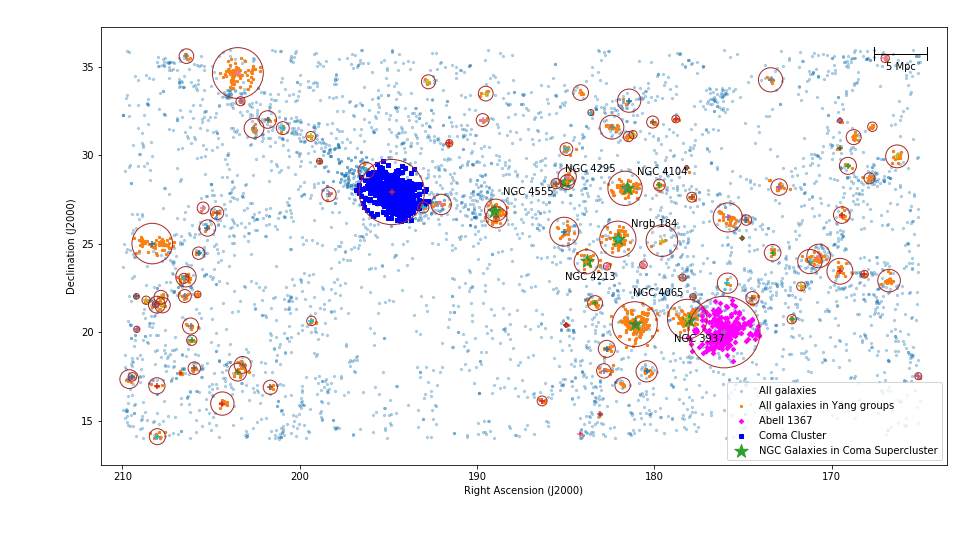}}
\caption{The Coma Supercluster, consisting of two Abell Clusters, A1656 (Coma cluster, Blue: Top Left) and A1367 (Magenta: Bottom Right), connected via a system of filaments, in which galaxies (blue) and galaxy groups (orange) are embedded.  Some of the important groups are labelled (mostly on the filament connecting the two clusters). The ``virial radius" of the circle around each groups is defined in
\S{\ref{subsection:infall}}.}
\label{sr-fig1}
\end{figure*}
\begin{table*}
\hspace*{-2cm}
\caption{The two Abell clusters within the Coma Supercluster from the Yang et al. Catalogue, (see \S\ref{section:Section-2} \& \S\ref{subsection:group} for more details). Properties are taken from the Yang et al. catalogue.}
\label{tab:Tab-1}
\begin{threeparttable}
\begin{tabular}{lcccccccccr}
\hline
Cluster & Group\_id & RA & Dec & Redshift & N & Group\_L\_{-19.5} & Stellar mass & Halo mass \\
&  & &  &($\rm km~s^{-1}$)& &$L_{*}$ &\M{} &\M{} \\ 
\hline
Abell 1656 (Coma) & 1 & 12 59 48.7 & +27 58 50 & 6950 & 641 & 12.2 & 12.6 & 14.8 \\
Abell 1367 & 5 & 11 44 44.6& +19 41 59 & 6450 & 272 & 11.9 & 12.3 & 14.5 \\
\hline
\end{tabular}
\end{threeparttable}
\end{table*}

\begin{table*}
\hspace*{-2cm}
\caption{Galaxy Groups within the Coma Supercluster (excluding Coma and Abell 1367) from the Yang et al. Catalogue, (see \S\ref{section:Section-2} \& \S\ref{subsection:group} for more details). The Group properties have been taken from the Yang et al. catalogue. The table in full form is available online.}
\label{tab:Tab-2}
\begin{threeparttable}
\begin{tabular}{lccccccccr}
\hline
Group\_id & RA & Dec & Redshift & N & Group\_L\_{-19.5} & Stellar mass & Halo mass   \\
&   &  &($\rm km~s^{-1}$)& &$L_{*}$ &\M{} &\M{} \\ 
\hline
14 & 12:04:26 & +20:27:43 & 7075 & 160 & 11.7 & 12.1 & 14.3 \\
68 & 13:33:57 & +34:39:13 & 7441 & 73 & 11.3 & 11.8 & 13.9 \\
71 & 12:06:37 & +28:08:20 & 8419 & 71 & 11.3 & 11.7 & 13.9 \\ 
72 & 11:52:35 & +20:45:15 & 6678 & 70 & 11.3 & 11.7 & 13.9 \\
77 & 12:08:14 & +25:15:52 & 6779 & 67 & 11.2 & 11.6 & 13.8 \\
83 & 13:53:15 & +25:00:59 & 8976 & 66 & 11.5 & 11.9 & 14.1 \\
235 & 12:15:21 & +23:59:22 & 6746 & 30 & 10.9 & 11.4 & 13.5 \\
322 & 12:35:47 & +26:54:08 & 7522 & 29 & 11.0 & 11.5 & 13.6\\
367 & 13:45:39 & +23:08:34 & 8975 & 27 & 11.0 & 11.4 & 13.6 \\
396 & 11:43:27 & +26:29:39 & 9092 & 25 & 11.2 & 11.6 & 13.8 \\
411 & 12:35:38 & +26:30:00 & 6601 & 26 & 10.8 & 11.2 & 13.2 \\
527 & 11:22:57 & +24:19:03 & 7641 & 22 & 10.7 & 11.2 & 13.1 \\
\hline
\end{tabular}
\end{threeparttable}
\end{table*}

\section{Characterising Environment in the Coma Supercluster}
\label{section:Section-3}

\subsection{Local and Global Environment in the Supercluster}

Fig.~\ref{sr-fig1} shows the spatial distribution of galaxies in the SDSS spectroscopic catalogue in the Coma Supercluster. The galaxies present in the Yang et al. group catalogue are shown in different colour.
Prominent groups are also marked along with the two major clusters, Coma and Abell 1367 constituting the Coma Supercluster.
Galaxies in the Coma supercluster are distributed on a web of filemants, with the two clusters forming prominent nodes. 
The clustering of galaxies highlights the distribution of the underlying dark matter potential.
 
Hence there is a significant variation in the density of galaxies confined per Mpc$^{2}$ from where the clusters are located, to the periphery of the supercluster. However this density variation is not isotropic. Instead we observe a higher gradient in the density variation along the direction of the spine of the filament joining the two clusters, compared to the other directions. 

To quantify this density variation in two dimensions, projected along the redshift space, and to identify the main spine of the web (in 1-D), the direction of maximum variation in the population density, we smooth the distribution using a kernel density estimator (KDE) algorithm \citep{Silverman86}. \somaked{This is a non-parametric technique to calculate a probability density estimator, represented by
\begin{equation}
\hat{f}_h(x)=\frac{1}{n}\sum_{i=1}^n K_h (x - x_i) = \frac{1}{nh} \sum_{i=1}^n K\Big(\frac{x-x_i}{h}\Big),
\label{eq:kernel}
\end{equation}
where $\hat{f}_h(x)$ is the returned kernel density estimate, $K_h(x)$ is the scaled kernel given by  $1/h K(x/h)$, $n$ is the number of data points in our sample and $h$ is the smoothing parameter, often known as the bandwidth. The kernel $K$ in our case is Gaussian, and the smoothing parameter (Scott's factor) $h\!=\!n^{-1/(d+4)}$, where $n$ is the number of galaxies in our case and $d=2$ is the number of dimensions.} The resulting smoothing kernel in our case has a Gaussian width of 0.22~deg.

This algorithm smooths over the filamentary-void network in the RA-Dec space, after projecting along the redshift axis
\footnote{The code is available at \url{https://docs.scipy.org/doc/scipy/reference/generated/scipy.stats.gaussian_kde.html}}.

The KDE-smoothed galaxy density distribution evaluated on a well-sampled grid is referred to as the overdensity or the local environment parameter of galaxies in the rest of the paper.

\begin{figure*}
\hspace*{-0.2cm}
\centerline{\includegraphics[width=0.9\paperwidth]{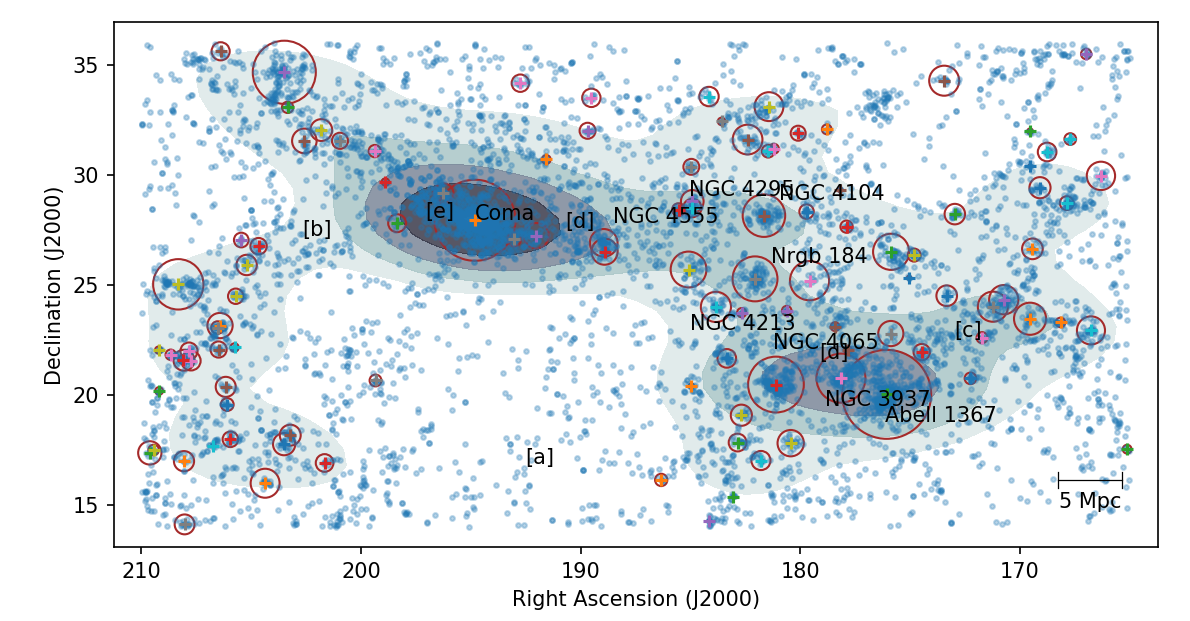}}
\caption{The Coma supercluster, consisting of the two rich clusters Coma (Top Left) and Abell 1367 (Bottom Right), along with groups of galaxies (prominent ones labelled) in the system. The radii of circles encircling groups and clusters is the virial radii defined in
\S{\ref{subsection:infall}}. 
The blue dots represent all the SDSS galaxies lying in the supercluster region. The contours represent the smoothed galaxy number density, estimated using kernel density estimator. The supercluster is divided into five main regions demarcated by the contours and quantifies galaxy environments based on mean galaxy number density (labelled : [a], [b], [c], [d], [e]). The [d] region appears twice and is present at the periphery of both clusters while [e] is present only at the boundary of the Coma cluster.}
\label{sr-fig2}
\end{figure*}

Although this algorithm characterises the local projected galaxy number density, which is similar to defining a $\Sigma_{5}$-like environment parameter, defined as the galaxy number density within the distance to the fifth nearest-neighbour and a velocity range of $\pm$1000 km~s$^{-1}$, but it is in principle a better estimate of the environment as it is relatively insensitive to Poisson errors. (Note that the redshift slice used in this paper is wider than $\pm$1000 km~s$^{-1}$.).
However the KDE estimate of the environment is quite different than $\Sigma_5$ and does not show a tight correlation. 
To characterise the local environment in terms of the number density of galaxies, these grid cells were combined into five bins, covering the whole range of overdensities.  These bins are shown as contours in Fig.~\ref{sr-fig2}. Contour levels represent mean overdensities, labelled [a], [b], [c], [d], [e], in increasing order from the outer-most edge to the interior. The contour levels sample the overdensities on a logarithmic scale spanning the entire dynamic range. 
The contour levels for the mean number of galaxies $\rm{Mpc^{-2}}$ (after the KDE algorithm is applied) are given in Table~\ref{tab:Tab-3}. 

The categories of  environment parameter [b]--[d] covers the `central filament' joining the two clusters, with the [d] environment, representing  the cluster outskirts, likely consists of a bundle of filaments entering the cluster. The category [e] encompasses galaxies in the core of Coma cluster,  accompanied by a few in-falling galaxies.

 Each galaxy in our sample was further assigned to  two classes: groups and non-groups (as defined in \S\ref{section:Section-2}), each group being assigned to the contour level corresponding to the group's centroid. Thus, in total we get ten environment bins: [a]--[e] according to smoothed density, each category split into group/non-group galaxies.

\begin{table*}
\hspace*{-2cm}
\caption{Environment parameters in the Coma Supercluster}

\begin{threeparttable}
\label{tab:Tab-3}
\begin{tabular}{lccccc}
\hline
Environment & $\sigma$ \tnote{$^\dag$} & \% of Galaxy Population & \multicolumn{3}{c}{Galaxy Number Density (N~$\rm{Mpc^{-2}}$)}\\
\hline
&&(Cumulative)& All & ~Coma & Abell 1367 \\
\hline
a & < 3.0 & 100 & 1.6 & ... & ... \\
b & 3.0 & 75 & 3.6 & ... & ...  \\
c & 5.2 & 45 & 5.2 & ... & ...   \\ 
d & 9.0 & 28 & 9.9 & 7.5 & 15.5  \\
e & 15.7 & 17 & 31.7 & 31.7 & ...  \\
\hline
\end{tabular}

\begin{tablenotes}\footnotesize
\item[$^\dag$] $\sigma$ = 0.00025, encloses 99.3\% galaxy population of the supercluster.
\end{tablenotes}
\end{threeparttable}
\end{table*}
\section{Results}
\label{section:Section-4}

\subsection{Pre-processing in groups across the large-scale environment}
\label{subsection:preprocess}
The phenomenon of `pre-processing' in groups, \citep{Cybulski2014, Mahajan2013, Hess2013, Wetzel2012, Fujita2004} was primarily invoked to justify the overabundance of red-sequence galaxies on the outskirts of clusters. The environment of a cluster, despite its deep dark matter potential well, is largely ineffective in enhancing quenching of star formation on such large scales. Thus a scenario, in which an population of galaxies, which have already been processed in environments such as groups or sub-clusters, while on the filaments of the cosmic web, before they 
being accreted onto the clusters, was favoured. This pre-processing in smaller structures might involve strangulation of the cold gas by the intra-group medium, or enhanced tidal interaction among members, or, in rare cases, ram pressure stripping of the fuel of star formation in galactic disks. Several pieces of evidence confirm this scenario, including morphological transformation of galaxies prior to cluster assembly \cite{Kautsch2008}, post-starburst galaxies present in  relatively high galaxy density regions in clusters  \citep{Mahajan2013} or the lack of neutral hydrogen (HI) detected galaxies in groups, and even less in high richness groups \citep{Hess2013}.
We study both the fraction of star-forming galaxies as well as the late type morphology fraction to understand the pre-processing taking place in groups. We compare the population inside the groups (four or more galaxies), to the population outside the groups across the five local environments.

\subsubsection{Star formation}
To calculate the star-forming fraction (hereafter; SF fraction), we consider all galaxies which have a specific star formation rate, SFR M$_{*}^{-1}$ (yr$^{-1}$) lying between $10^{-8}$ to $10^{-11}$ as star-forming. While calculating the fraction of star-forming galaxies, we divide their number by the whole population in the given environment. (As mentioned before, we treat all missing galaxies in the \cite{Brinchmann2004} catalogue, with no match in our catalogue, as non-star-forming, since the reason for their non-inclusion is a failed fit, which is most likely due to a poor signal to noise ratio in both continuum and absorption lines, indicative of low star-formation). 
Fig.~\ref{sr-fig3} shows the fraction of star-forming galaxies in groups and non-groups binned in their respective local environments. The error bars adopted here are the binomial confidence intervals, calculated using the binomial CI formula \citet{Cameron2011} at the 1$\sigma$ confidence level. 

The pre-processing in galaxies present in groups is evident across all environments in Fig.~\ref{sr-fig3}. In a given range of over-densities (represented by the region lying between contours), galaxies emerging from groups have a higher quenched fraction as compared to the field counterpart. The star-formation fraction originating from group differs from the non-groups.A Kolmogorov-Smirnov (K-S) test performed between the group and non-group subsamples for environments [a]--[e] shows that the probability of the pairs being drawn from the same underlying distribution is low (0.20), indicating that they have been sampled from moderately different populations. Thus we can conclude that pre-processing is prevalent in the groups across the entire supercluster.

The trend in star-formation for non-group galaxies,  as a function of local density, is flat for the initial three environmental bins, [a], [b] and [c] but decreases significantly in the last two environment categories, [d] and [e].  In the case of galaxies in groups, we initially see a significant decline in the star-forming fraction of groups from [a] to [c]. However, the trend for groups in the [d] environment reverses, exhibiting an enhanced SF fraction for groups as compared to its previous environment. This sudden increase in the SF fraction in a denser environment is seen only for galaxies in groups and is not present in the galaxies which are not part of any group. 

However, this observation is an exception, as it does not quite follow the gradual decline of star-formation properties as the local galaxy density increases, from values typical of the field, gradually to values typical of the outskirts and cores of clusters \citep{Dressler1980, Kauffmann2004, Balogh2005}. The overall trend in the SF fraction against the local environment, is also consistent with the \citep{Dressler1980} picture, where an increase in local over-density would mean a higher degree of quenching.

One can argue that the enhanced star formation in environment [d] could be due to the underlying galaxy population being rich in dwarf galaxies, which are known to form more stars as compared to giants at the current epoch. The observed trend could then be a mere manifestation of a population bias in the sample. To examine this possibility, we further divide our sample into faint and the bright galaxies, with luminosity being a proxy for their mass. We have chosen this boundary at $M_{r}$ = -18.5, which approximately corresponds to the absolute magnitude of the Large Magellanic Cloud (LMC) in the $r$-band.

Fig.~\ref{sr-fig4} shows the trend in groups, which is split into the brighter population ($M_{r} < -18.5$) and fainter population ($M_{r} > -18.5$), represented by the blue circles in each of the sub-plots. The fainter galaxies both in groups and non-groups show a higher star-forming fraction than their more massive counterparts.

A K-S test performed between the fraction of star-forming bright galaxies present in and out of groups returned a $p$-value of 0.20, supporting the hypothesis that these two distributions coming from groups and outside groups are drawn from moderately different populations. The same exercise performed for the faint galaxies returned a $p$-value of 0.036, implying significantly different parent populations in and out of groups. This seems to indicate that the phenomenon of pre-processing in the group environment is more effective for the fainter population than in their brighter counterparts. Such systems are largely quenched in both the environments.
This implies that the fainter galaxies are perturbed more easily by the group environment, which is consistent with the earlier studies \citep[e.g][]{Mahajan2011}.

On the other hand, a K-S test performed for the fraction of star-forming bright galaxies lying in groups to the faint galaxies lying in groups returned a low $p$-value (0.036), implying that the bright and faint galaxies in groups are drawn from different underlying populations. 
As for the non-group sample, the K-S test performed with the fraction of star-forming galaxies, between the bright and faint samples, yields a very low $p$-value (0.003), which suggests that the fainter galaxies can be statistically distinguishable from their brighter counterparts in terms of their star-formation activity, irrespective of whether they form a part of the group environment or not.

Thus we can identify the luminosity or mass of the galaxies as one of the factors in their evolution.

We conclude that isolated galaxies form more stars than the ones in groups in spite of the underlying local densities being comparable for both the sub-samples \citep{porter2007, porter2008, Haines2018}. The trend across the environments also persists with the earlier observed trend for both groups and non-groups. For the group environment, even after the split based on the luminosity of the galaxies, we continue to see a monotonous decline in SF fraction from [a] to [c] followed by an enhancement in [d]. 
It is possible that the enhancement in star formation seen in groups in [d], could be due to the infall regions of the rich cluster, Abell~1367 in this case. The rise in star-formation rate immediately outside the virial radii of clusters that are fed by supercluster filaments, over and above the general trend of quenched star-formation with decreasing distance from cluster centres, has been reported in samples of galaxies in superclusters \citep{porter2008, Mahajan2012, Haines2018} had shown a similar trend, where the galaxies, just before their infall into clusters, showed an enhancement in their star-formation properties. In our case, we see the non-group galaxies still hold a higher star-forming fraction than the corresponding galaxies belonging to groups in this bin; the individual galaxies showed no enhancement before being assimilated in the cluster. If the above scenario is true, it would imply that the cluster has a pronounced effect, if at all, only on galaxies in groups. Thus, contrary to what is found in the literature, this seems to be an exception where the magnitude of the galaxy magnitude and its over-density parameter (i.e. the local environment) are insufficient to predict the incidence of enhanced star-formation seen. Thus our priors could be misleading, and the larger picture of whether the galaxy is a part of a group or not becomes critical in determining the galaxy's characteristics. 

\subsubsection{Late-type morphological fraction}
It has long been known that the morphology of the galaxies in a particular volume  depends on the local environment \citep[e.g.][]{Dressler1980}. The group environment, with its low relative velocities of the member galaxies, encourages galaxy interactions, and consequently leads to an enhanced rate of morphological transformations \citep[e.g][]{Miles2004}. Here, to explore this connection, we have computed the fraction of late-types in each of our environment categories. All galaxies in our sample have been eyeballed for classification, using the full-resolution SDSS DR13 \& The Dark Energy Camera Legacy Survey (DECaLS, {\texttt{http://legacysurvey.org/decamls/}}) images. The galaxies were broadly classified into late-type (disk galaxies and irregulars) and early-type (ellipticals, S0s and dwarf spheroids). 

\begin{figure}
\hspace{-0.75cm}
\includegraphics[width = 0.55\textwidth]{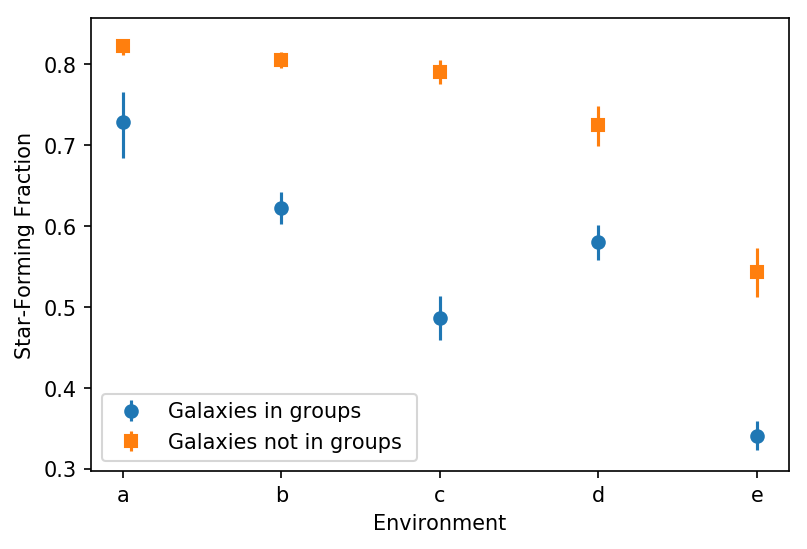}
\caption{Fraction of star-forming galaxies ($\log SFR/M^*\geq-11$ yr$^{-1}$) in each of the five environment bins. The blue circles represent galaxies in groups while the orange square represents the galaxies not belonging to any group. Note that the ``group" in [e] environment is the Coma cluster while the [d] environment consists of both Abell 1367 as well as the outer periphery of the Coma cluster (see Fig.~\ref{sr-fig2}). The adopted error bar represents the binomial confidence interval. We find the overall fraction of star-forming galaxies is high if they do not belong to any group. Overall there is a decline in the content of star-forming activity as we advance towards the higher density environment bins and is seen for both the categories. The only exception is seen in the group galaxies belonging to the [d] environment bin, departing from the overall decline. We discuss further the caveats seen in the trends in \S\ref{subsection:Impact of galaxies in groups in the infall region}}
\label{sr-fig3}
\end{figure}

\begin{figure*}
\hspace*{-1cm}
\includegraphics[scale = 0.65]{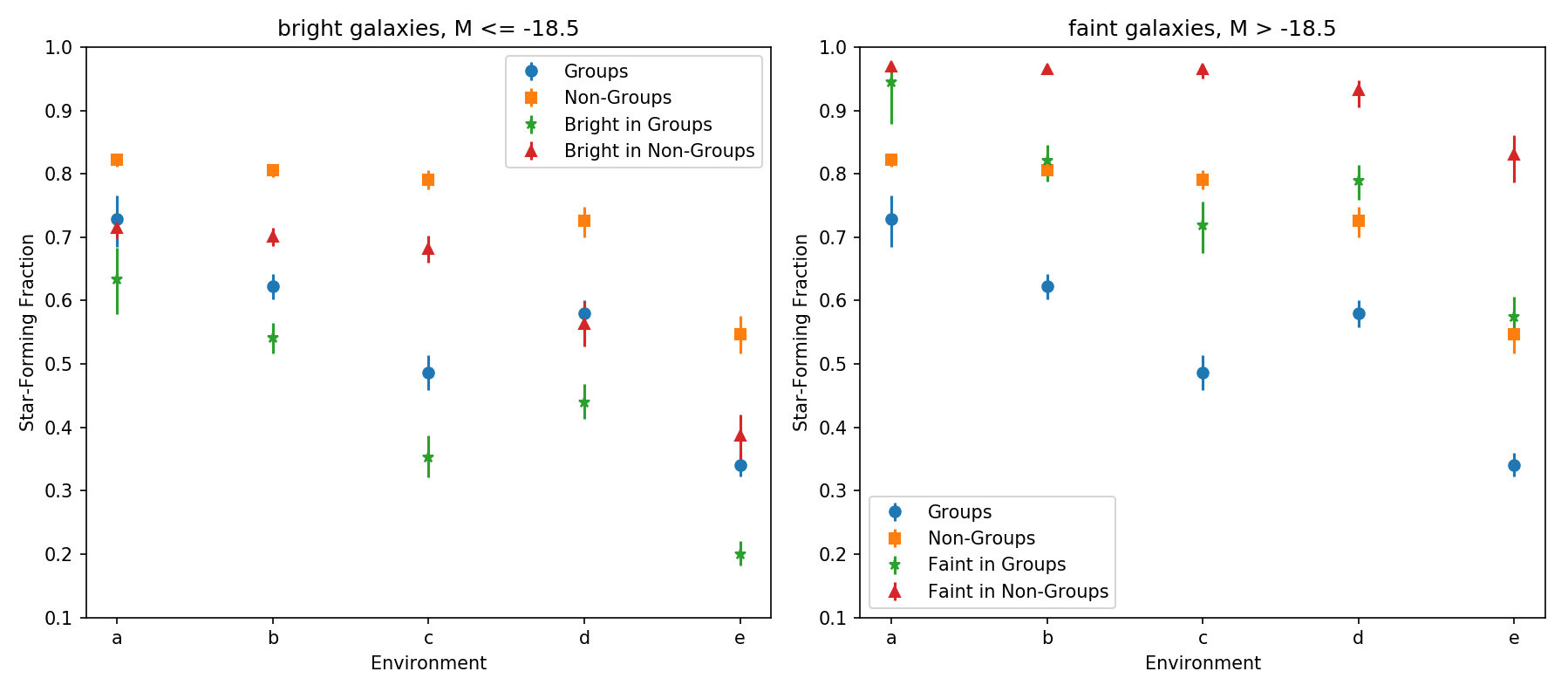}
\caption{Fraction of star-forming galaxies in the five environment bins. Galaxies have been divided into two classes (Left and Right figures) based on their absolute magnitudes (greater or less than $M_R\!=\!-18.5$). Non-group galaxies (orange squares) and galaxies in groups (blue circles) are plotted in both figures for comparison. We find the trends in group and non-group galaxies are similar even when they were split by their magnitudes. The only appreciable difference comes in the overall fraction of SF galaxies: bright (lower) and faint (higher), compared to when all of them were combined, as shown previously in Fig~\ref{sr-fig3}.}
\label{sr-fig4}
\end{figure*}

\begin{figure}
\hspace*{-0.5cm}
\includegraphics[width=0.5\textwidth]{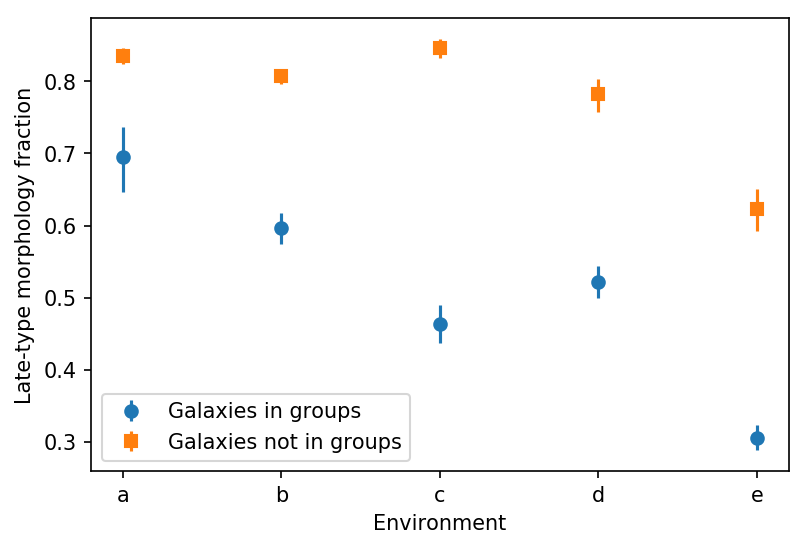}
\caption{Same as Fig.~\ref{sr-fig3}, but for the late-type morphology. The trend observed in this case turns out to be quite similar to the one seen for star-formation, especially for the galaxies in the group with a declining trend across [a--e] and slight enhancement present in the [d] environmental bin and higher fraction late-type galaxies among the non-group galaxies instead of the group members.}
\label{sr-fig5}
\end{figure}

In Fig.~\ref{sr-fig5}, the late-type fraction is substantially lower in groups compared to that in  non-groups, which is  supporting evidence for enhanced pre-processing in the environment of groups. The trend of the late-type morphology in groups is similar to the one seen in the case of star-formation in Fig.~\ref{sr-fig3}, with exactly the same exception of an enhanced fraction of late-type morphology in the groups lying in the [d] environment. The groups in [e], i.e. the Coma cluster has the least amount of late-types. The late-type fraction occurring in the Coma cluster lie in the range obtained in earlier studies and is about 30\% within a degree from the centre \citep{Bernstein1994,Doi1995,Beijersbergen2003}. In case of isolated galaxies too we find similar trend persists between late-type and SF fraction across all environmental bins conforming morphology and star-formation of the galaxies could be closely related \citep[see, e.g][]{Bait2017}. K-S test performed on the two sets returned a low p-value (0.0037), confirming the two distributions of late-type fraction emerging from the group and non-group galaxies were statistically distinct in their morphology.

\subsection{Role of the group and local environments in nuclear activity}

\subsubsection{Optical AGN across the environment categories}
A key ingredient in the evolution of galaxies is the evolution of their supermassive black holes (SMBHs), which primarily grow through minor mergers and accretion of gas, and often act as active galactic nuclei (AGNs), e.g. \cite{Hickox2009, Hopkins2010, Steinborn2018}. The connection between galaxies and SMBHs is indicated by the observed tight correlation between the masses of the central SMBH and that of galaxy bulges \citep[e.g.][]{Hopkins2006}, or proxies such that velocity dispersion \citep[e.g.][]{Richstone1998}. The likelihood of the host galaxies possessing a bulge increases if it is a member of a group than otherwise in a field. Thus we can expect that the groups are a more favourable site to harbour AGNs.

\begin{figure}
\hspace{-0.5cm}
\includegraphics[width=0.5\textwidth]{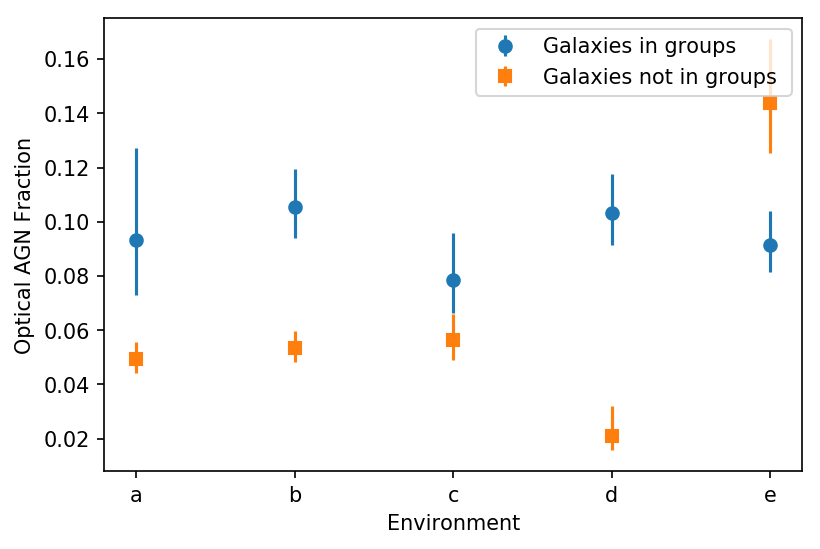}
\caption{Same as Fig~\ref{sr-fig3}, but for optical AGN. Group galaxies have a higher fraction of optical AGNs as compared to the non-group galaxies in the first four environmental bins. The non-group galaxies are significantly high on the AGN content in the [e] environmental bin compared to both the groups in [e] as well as other non-group galaxies in the preceding environmental bins. The overall trend with respect to the environment is flat in case of group-galaxies hosting an AGN.}
\label{sr-fig6}
\end{figure}

\somaked{We see in Fig.~\ref{sr-fig6}, the AGN fraction is similar across all environments [a]--[c], which lie outside the clusters, at a level of $\sim$5\% in non-group galaxies, but enhanced to about $\sim$10\% in group galaxies. The two clusters have similar values as that found in the groups. This is not true in the category [e] where the non-group galaxies are enhanced to 14\%. This enhanced fraction was seen in a previous study by \citet{Gavazzi2011} of the same Supercluster, who had found that the number density of optical AGN hosts anti-correlates with the local density of galaxies, observing that the frequency of AGNs drop by a factor of two from the outskirts of the Coma cluster to the core. Various studies \citep[including][]{Shen2007,Manzer2014} 
have found a significantly higher fraction of AGNs (by as much as 25\%) in groups, compared to isolated galaxies. One would expect that galaxy groups are usually associated with a higher fraction of AGNs since tidal interactions between major galaxies are more common in groups, due to low relative velocities.}

In these groups and clusters, apart from the brightest group/cluster galaxies, optical AGNs are more likely to be found in the outskirts than in the cores. However, for our limited sample we found the population of optical AGNs to be clustered around the middle of groups.
The K-S test resulted in a p-value of 0.0361, implying that two populations emerging from groups and otherwise were likely distinct in terms of their fractional AGN activity.

\subsubsection{Radio AGN across the environment categories}

The incidence of radio AGNs in normal galaxies rises towards the luminous end of the red sequence, indicating that they reside in massive early-type galaxies, with fewer sources detected in the blue cloud \cite[e.g.][]{Hickox2009}. Since Radio AGNs, unlike optical AGNs, usually accrete in a radiatively inefficient manner, and thus they do not require a cold gas supply.  The physical mechanisms that quench cold gas accretion in massive groups and clusters are not effective in quenching radio AGNs. Since early-type galaxies mostly reside in massive groups and clusters, we have a high probability of finding radio-active AGNs in groups, at least in their virialised cores.

\begin{figure}
\hspace{-0.5cm}
\includegraphics[width = 0.5\textwidth]{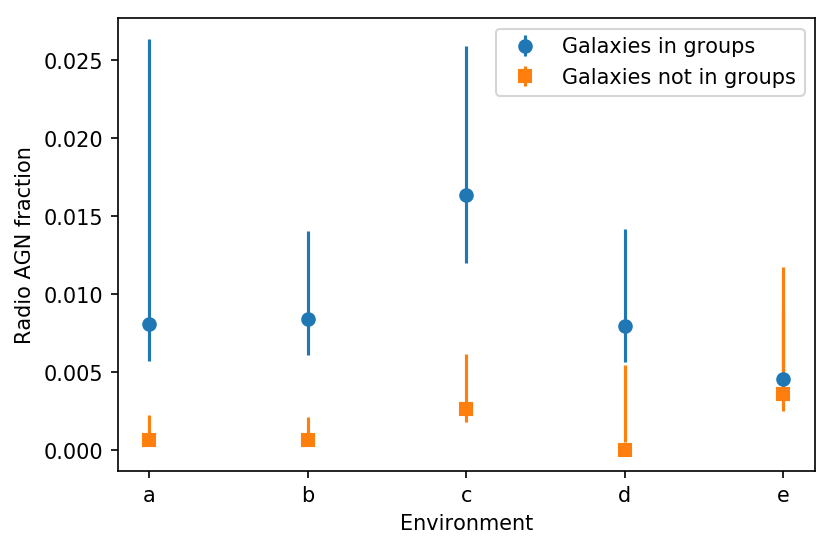}
\caption{Same as Fig~\ref{sr-fig3}, but for fraction of galaxies hosting a radio AGN. We find significantly higher fraction for galaxies in groups than in non-group galaxies, except for the last environmental bin (Coma cluster) where both are equal. Both galaxies in groups and fields exhibit a flat trend in the fraction of radio AGN across the environment bins.}
\label{sr-fig7}
\end{figure}

We looked at the distribution of radio AGNs across our environments, which have been separated out by \citet{Best2012} (see \S\ref{subsection:radio AGN} for details). However, the total number of radio AGNs in our sample (24) is too small to establish a trend across the large-scale environment. Out of these 24, 15 lie in groups, 5 are associated with non-group galaxies, the remaining 4 being in clusters. In Fig.~\ref{sr-fig7}, we show the fraction of all galaxies hosting a radio AGN from \cite{Best2012}, in the respective local environmental bins.
The fraction of radio AGNs is roughly similar across all environments [a]--[e] , at a level of $\sim$0.15\% in non-group galaxies, but enhanced to about $\sim$0.7\% in group galaxies (as seen earlier in \citet{Martini2006}).
This is consistent with \citet{Best2012} who for a substantially larger volume found that galaxies present in groups and clusters, mostly surrounded by the hot X-ray haloes, were likely to show a higher presence of radiatively inefficient accretion AGNs (LERGs). In an earlier work too, \cite{Best2005} had shown the host galaxies of the radio-loud AGNs to be larger in size, inhabiting richer environments than other galaxies. 

The radial distribution of radio galaxies for all groups stacked together show an enhancement in the cores, decreasing outwards. The K-S test returned a low p-value (0.004) confirming the population of radio galaxies is distinct statistically for group and non-group and emerged from distinct parent sample. 

\subsection{Influence of Group Halo Mass on Galaxy Evolution}
Several studies have investigated \citep[e.g.][]{Wetzel2012,Woo2013,Roberts2016,Roberts2017,Hess2013} the influence of the mass of the dark halo of a group (``halo mass") or the richness of a group (number of galaxies) on its quenching efficiency. There is overall agreement that the degree of pre-processing in groups increases with an increase in the group halo mass. The most vulnerable systems are the dwarf satellite galaxies, which, due to their shallow gravitational potential, can lose their gas in relative readiness and appear in the category of galaxies with quenched star formation. 

By stacking groups in at least four richness bins, \citet{Hess2013} showed that, as the group richness increases, the fraction of HI detected galaxies becomes substantially low. Galaxies that have counterpart in the ALFALFA survey (of HI in galaxies) tend to occupy the outskirts of rich groups. They suggest that as the groups grow richer, the gas in their core is gradually depleted, and is regulated by fresh supply of fuel only from the surrounding cosmic web \citep[e.g.][]{Kleiner2017}. Thus establishing the importance of halo properties
responsible for activity in their member galaxies. We look at the impact of halo mass (richness of the group) on galaxies physical properties.
\begin{figure}
\hspace{-0.5cm}
\includegraphics[width=0.5\textwidth]{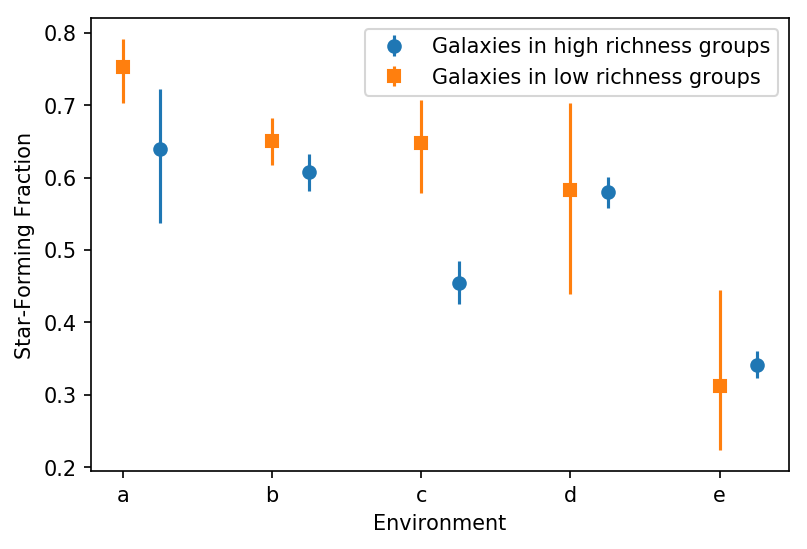}
\caption{Fraction of star-forming galaxies present in high richness ($N\geq 10$) groups (blue circles) and low richness ($N < 10$) groups (orange squares). Star-forming activity is observed to be alike for low and high richness groups, differing only in the [c] bin. The overall activity of galaxies in groups decline as we move to a higher density environment with the high richness groups throwing an exception in the [d] bin. Both these observations are consistent with our earlier picture (Fig.~\ref{sr-fig3}). The [e] bin mainly consisting the Coma cluster is largely quenched.}
\label{sr-fig8}
\end{figure}

\subsubsection{Star formation and halo mass}

So far we have assumed that the richness of the group traces the underlying dark matter halo, and thus it was used as a proxy for the halo mass. In Fig.~\ref{sr-fig8}, we observe the trend of the fraction of star-forming galaxies, across the local environment, in two richness bins high (low), corresponding to more (less) than ten galaxy members. Here we observe that the low mass haloes portray a rather flat trend for [a]--[d] followed by a sudden decline corresponding to the densest environmental bin. In case of high mass haloes, we observe a consistent decline in the overall trend from [b] to [c], while [d] rises again, thus the trend looks familiar like Fig.~\ref{sr-fig3}. On the other hand, the low mass haloes remain largely unaffected by their local environment.

\begin{figure}
\hspace{-0.5cm}
\includegraphics[width=1.1\columnwidth]{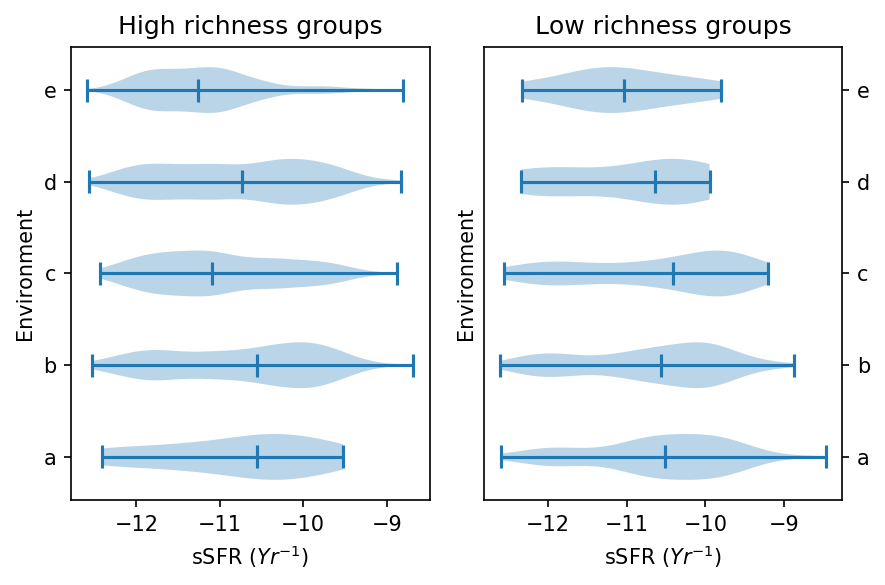}
\caption{
{\it A Violin plot gives the distribution of the data along with its probability density as a means of data visualisation.} Here the plot shows the bimodal sSFR distribution in galaxy populations, representing
the distribution of galaxies, in bins of sSFR, $\rm{log}(SFR/M_{*})$ in high-richness groups (left) and low-richness groups (right) across the five local environments. For the low-richness groups the population gets skewed towards the quenched population ($\rm{log}(SFR/M_{*}) < -11$) as we go from [c] to [e]. In the case of high richness groups, however, we observe a similar trend but in the main star-forming sequence ($\rm{log}(SFR/M_{*}) > -11 $) being depleted as we go from [b] to [c], but it unexpectedly shows an enhancement in the [d] environment, quenching completely in the [e] bin.}
\label{sr-fig9}
\end{figure}

Fig.~\ref{sr-fig9} represents the distribution of sSFR (yr$^{-1}$) for all galaxies, in high (low) richness group in the left (right) panel for each of the five environmental bins, in the form of so-called {\it violin plots}. 
 
The violin plot shows the probability density of data, often smoothed using a kernel density estimator. In addition, it demarcates several intervals, for example, median (or mean), quarter percentiles and extrema, and, as a result, quantifies data across one (or more) categorical variables, simplifying the comparison of distributions. Here the violin plot highlights the bi-modality seen in the distribution of galaxy population in the values of their sSFR. The bi-modal population consists of the star-forming main sequence (sSFR $< -11$~yr$^{-1}$) and the quenched sequence (sSFR $> -11$~yr$^{-1}$).

As we go from the bottom panel to the top panel, advancing towards the higher densities we would expect the distribution of galaxies to shift and change accordingly on the sSFR scale. 

In the case of low-richness groups, we observe a consistent enhancement of the quenched sequence as we move upwards, from  [c] to [e] (the first two environmental bins do not show any significant trend). In the case of high-richness groups, however, we observe, from [b] to [c], there is a dip in the population of star-forming main sequence. However, in the transition from [c] to [d], the population comprising of the main star-forming sequence increases, resulting in the diminishing of the quenched sequence. Thus the population present in the environment [d] is predominantly star-forming, even when it is a part of high-richness group bin. This contradicts the trend found earlier by \citet{Wetzel2012}, who found the population of satellites are more likely to be quenched with an increase in the halo mass of the group. The bin [e] (dominated by the Coma cluster) comprises mainly of the quenched population.

However, a K-S test performed on the two distributions of sSFR in high and low mass haloes for [a]--[e] returned a $p$-value of (0.2) implying that the two populations were moderately different and that the impact of the mass of a halo by itself is insufficient to determine the fate of the star-forming population contained within it.

\subsubsection{Late-type morphology and halo mass}

\begin{figure}
\hspace*{-0.5cm}
\includegraphics[width=0.5\textwidth]{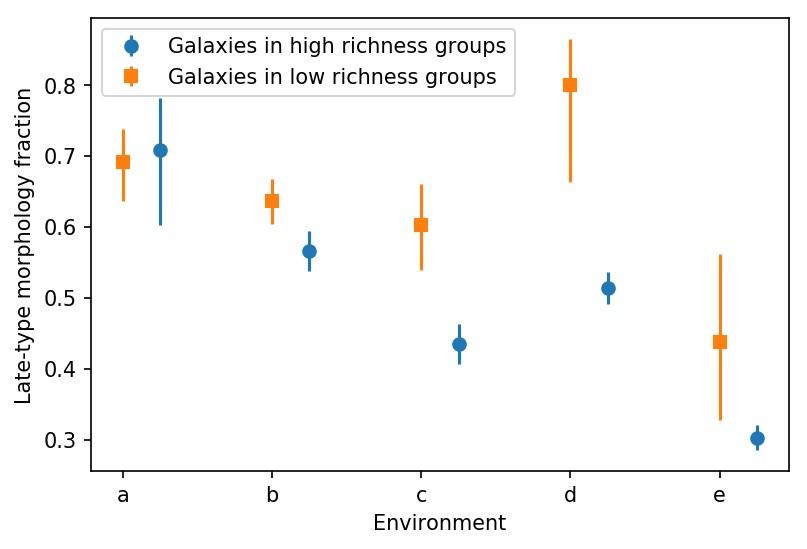}
\caption{This is similar to Fig.~\ref{sr-fig8}, but for late-type morphology. Fraction of late-type morphology exhibits a similar trend with environment as that of the fraction of star-forming galaxies. Overall the content of late-type morphology is higher in the low-mass haloes as compared to the high-mass ones. There is an overall declining trend from [a]--[e] in both high and low richness bins as is true for all groups in the sample (see Fig.~\ref{sr-fig5}).
The slight enhancement in the fraction of late-type galaxies seen earlier in Fig.~\ref{sr-fig8} is also reflected in the [d] bin albeit this time for both the richness categories.}
\label{sr-fig10}
\end{figure}

\somaked{In Fig.~\ref{sr-fig10}, we look at the trend of the fraction of galaxies with late-type morphology present in the high and low richness groups. The fraction of late-type galaxies has a similar trend compared to the fraction of star-forming galaxies, as seen in Fig.~\ref{sr-fig8}. Galaxies with late-type morphology are preferentially found in groups low in richness. Groups in the high richness regime showed a systematic decline in the fraction of late-type galaxies, while for low-richness groups the tend is flat between environment bins [a]-[c]. Thereafter, for both the halo richness regimes, this fraction rises in [d], followed by a final significant decline in [e]. The observed trend seen in the late-type fraction across the environment is largely consistent with a large number of studies conducted from large-scale samples in the vein of the \cite{Dressler1980} morphology-density relation. \citet{Roberts2016,Weinmann2006} have shown that the late-type fraction declines for satellite galaxies as the halo mass is increased, which are  consistent with our observed trend for the [b]--[e] environments. }

\somaked{The K-S test performed between the two samples returned a p-value of 0.2 implying that the two underlying populations were moderately distinct, considering our sample contains the high mass (central) galaxies too which are mostly the early-type and unaffected by the halo mass. Hence we can conclude that the group richness have a significant impact on galaxy morphology and that the late-type morphology increases with a decrease in the halo-mass supporting a higher degree of pre-processing prevalent in high richness groups.}

\subsubsection{Optical AGN and halo mass}

\begin{figure}
\hspace{-0.5cm}
\includegraphics[width=0.5\textwidth]{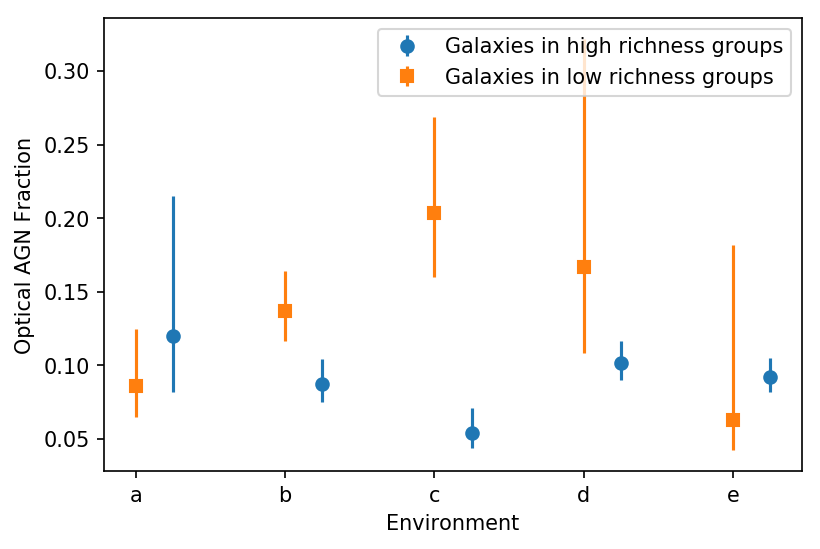}
\caption{Same as Fig.~\ref{sr-fig8}, but for optical AGNs. We see a significant presence of optical AGNs in groups of low richness as compared to their high richness counterparts present in the [b]--[d] environmental bins in the Coma Supercluster. We see an overall flat trend in the distribution of groups in both low and high richness bin, regarding its optical AGN content, lying in different local environmental zones.}
\label{sr-fig11}
\end{figure}

\somaked{
In Fig.~\ref{sr-fig11}, we observe that the middle three environments show a significant difference in the optical AGN fraction, between galaxies belonging belonging to low and high richness, with low-richness groups hosting larger fraction of optical AGN than high-richness groups. This is consistent with \cite{Gordon2018a}, who found that high mass groups support the inhibition of AGN in their cores, and that when groups are split into bins by mass, only the lowest quartile showed significant enhancement in the AGN content. In our case we have split our sample into two, by richness, but our observations agree with this trend.}

\somaked{The low-richness groups show a flat distribution across all environments, with a mean optical AGN content of 8\%. The high richness groups on the other hand vary between 6-20\%, with a maximum in [c]. This is consistent with \cite{Gavazzi2011} who observed that the frequency of AGNs decrease as we go towards the cluster centre in this particular supercluster. The decline in the number of AGNs could simply arise due to the underlying distribution of groups that are present in each of the environment bins, more numerous in the middle than at both ends, the central galaxies of these groups being more likely to harbour active nuclei.}

\begin{figure*}
\includegraphics[scale = 0.5]{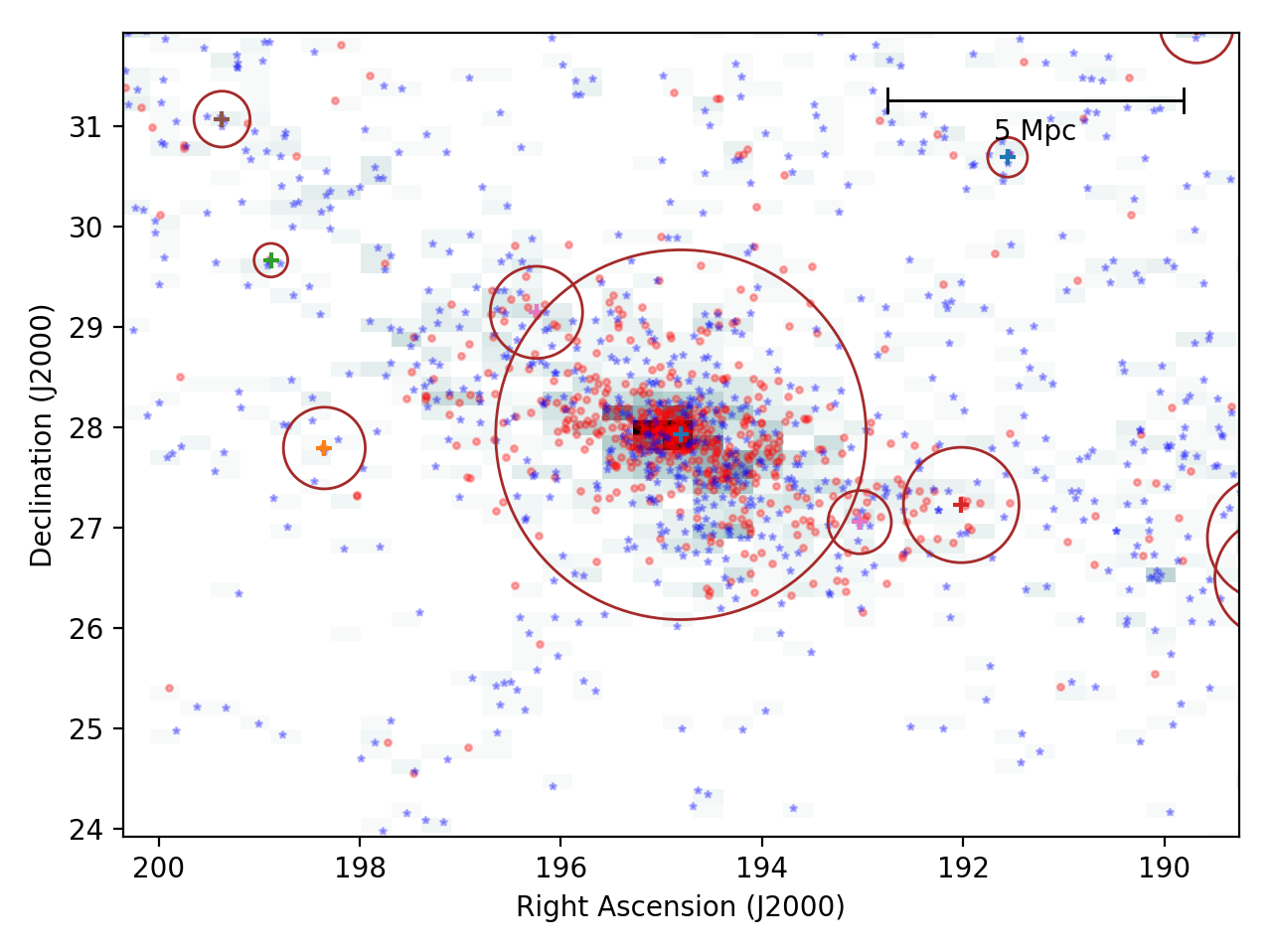}
\includegraphics[scale = 0.5]{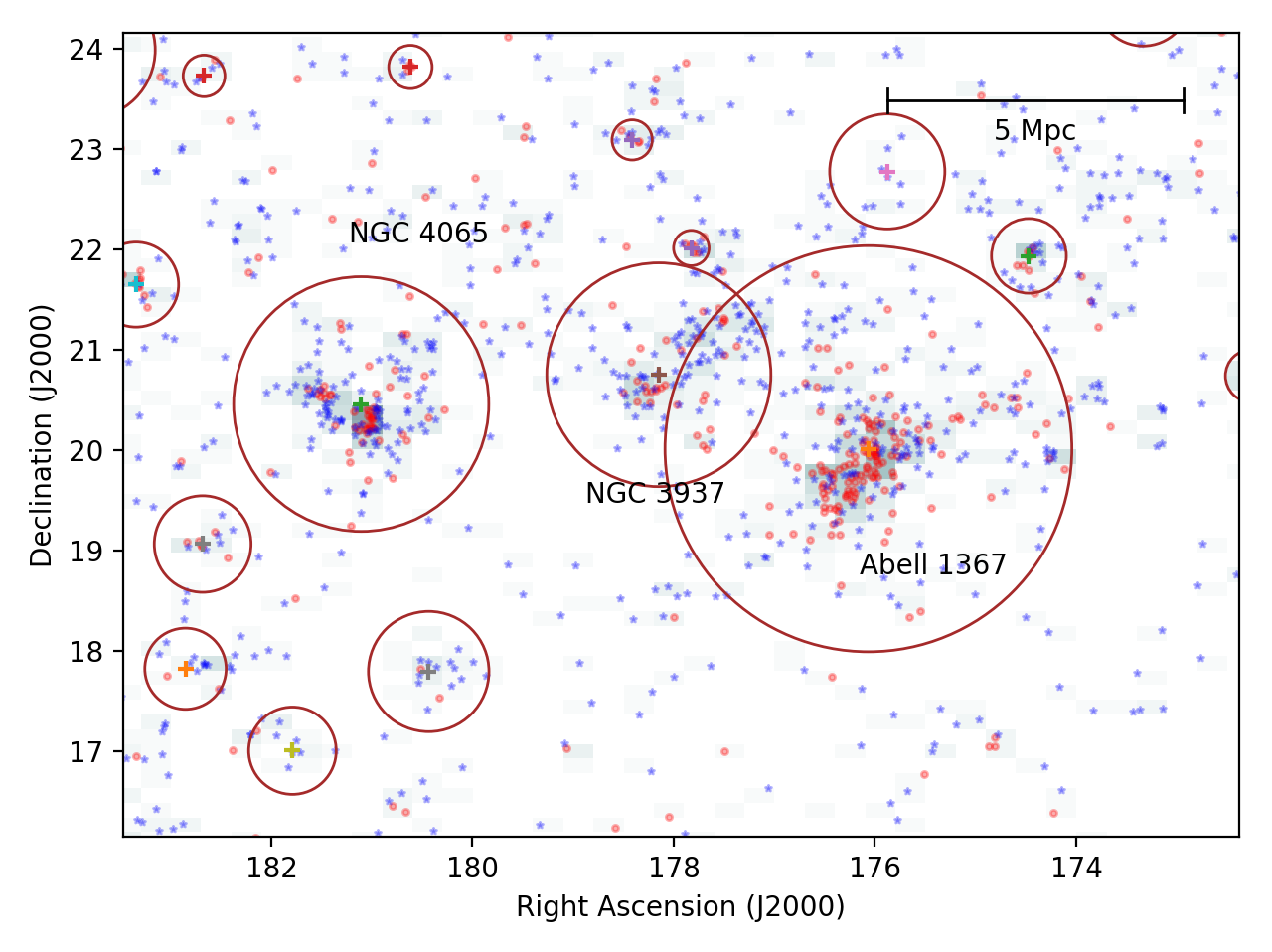}
\hspace*{-0.25cm}
\includegraphics[scale = 0.6]{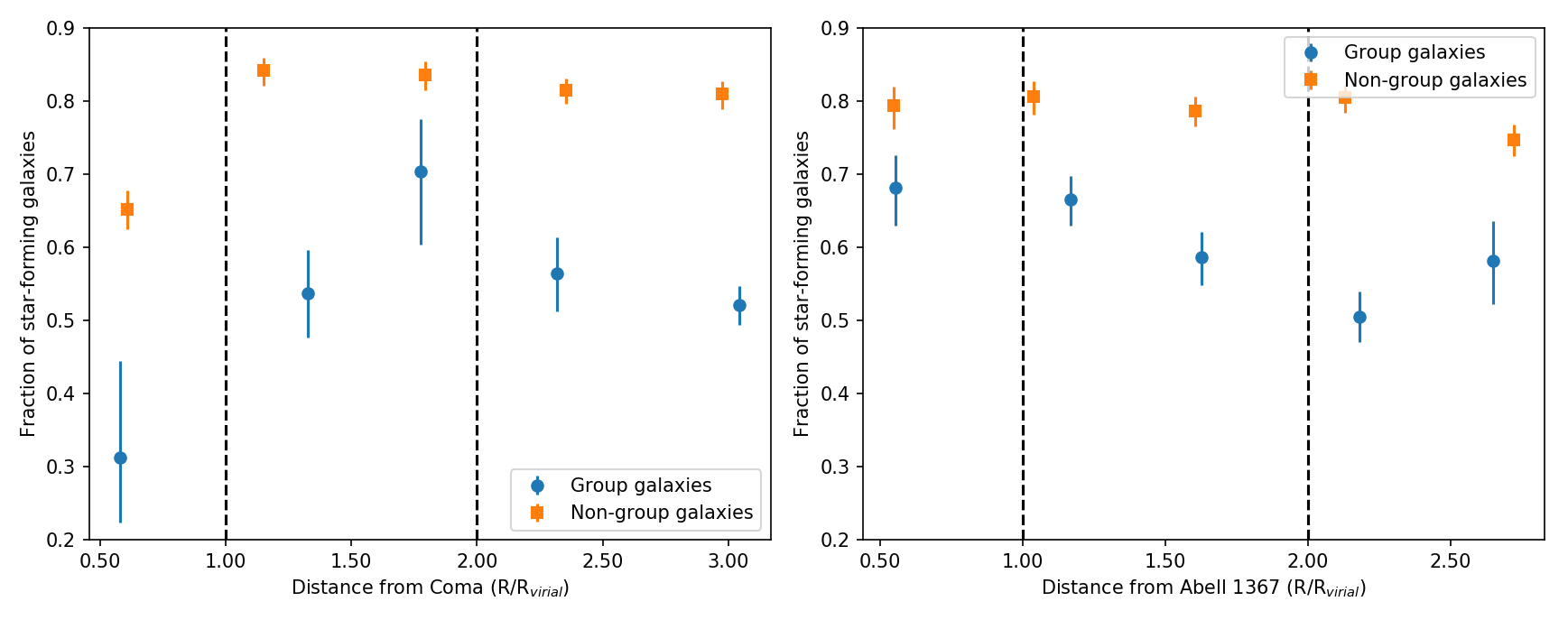}
\caption{The fraction of star-forming galaxies in the Coma cluster (left) and Abell 1367 (right), and their immediate surroundings in our sample over-plotted on the 2D histogram representing the number of galaxies in each bin. Top panel: Colours representing the sSFR of individual galaxies: blue representing those whose $sSFR \geq -11$ $\rm{yr}^{-1}$, which we have defined as our star-forming sample (see text), the rest being denoted red. Bottom panel: 1D profile for star-forming activity in galaxies both in groups and not in groups drawn for a projected radius extending up to six times R$_{\rm virial}$ of the respective clusters excluding the two clusters. R$_{\rm virial}$ for Coma and Abell 1367 are 1.84 and 2.02 Degrees respectively and are defined in \S{\ref{subsection:infall}}. The two-dashed lines represent R$_{\rm virial}$ and 2~R$_{\rm virial}$ of the respective clusters and the region bounded by the two lines comprise of the in-falling and back-splash population. The trend for the groups in the surroundings of the two clusters is striking while one increases as we move away from Coma, the trend is quite the opposite in the case of Abell 1367. The trend in case of non-group galaxies remain almost similar for both cases except for very near to Coma which shows a significant decline. The surroundings of Coma, including the two groups, are mostly red, while the scenario is quite the opposite in the surroundings of Abell 1367.}
\label{sr-fig12}
\end{figure*}

\begin{figure}
\hspace*{-0.5cm}
\includegraphics[scale=0.65]{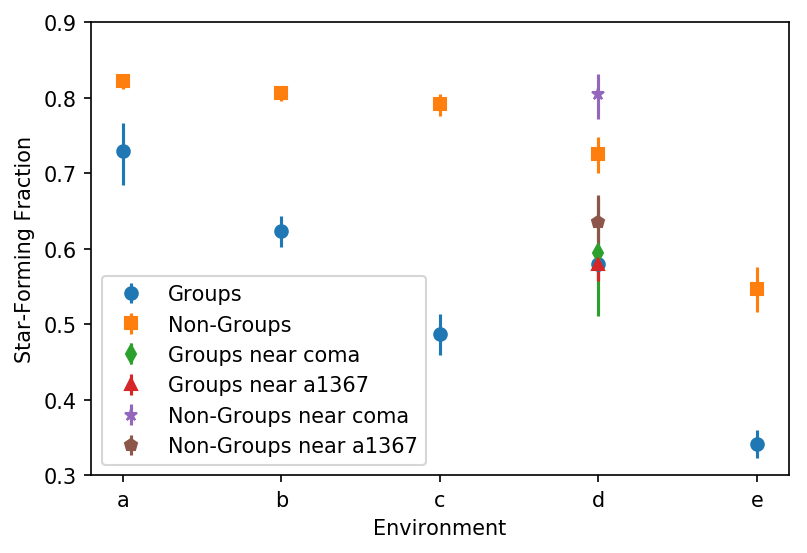}
\caption{The fraction of star-forming galaxies, both in groups and otherwise, drawn separately for the two contours enclosing Coma and Abell~1367. Plotted are the values of the star-forming fraction for all galaxies in groups (blue circles) and not in groups (orange squares) for comparison. The first three environments are common for both sets. The [d] environment bin has contribution from Abell~1367 and outskirts of Coma. Galaxies out of groups towards Coma display an enhanced star-formation compared to those near Abell~1367. The trend in groups, namely an expected decline in SF fraction for the first three environmental bins (arranged in increasing order of their mean over-densities), followed by observing an unexpected spike in the SF fraction in the [d] environmental bin has an equal contribution coming from the surrounding of both the clusters. Coma cluster, in the [e] environmental bin, is mostly quenched. The fraction of individual galaxies near Coma has an enhanced fraction of star-formation than Abell~1367.}
\label{sr-fig13}
\end{figure}

\begin{table*}
\label{tab:Tab-4}
  \centering
  
  \caption{Galaxy properties for galaxies in and out of groups in each of the five environmental bins}
\begin{tabular}{lrccccccccc}

\hline
 &  & Non-&  & Late-Type & Radio & Optical &  & & Radio & Optical \\
Env & Group & Group& SF & Group & AGN & AGN & SF & Late-Type & AGN & AGN \\
 & &   & Group &  &Group   &Group &Non-Group&Non-Group&Non-Group&Non-Group\\ 
\hline
\hline
\\
a  & 118   &  1457  & 73 & 111 & 1 & 11 &1198 & 982 &1 & 72\\
b  & 589  &   1550  & 312 & 376 & 5 & 62 &1249 & 1201 & 1 & 83\\
c  & 331  &   747  & 170 & 171 & 6 & 26 &591 & 631 & 2 & 42\\
d  & 543   &  335  & 262 & 234 & 4 & 56 &243 & 262 & 0 & 7\\
e  & 657   &  278  & 200 & 204 & 3 & 60 &151 & 173 & 1 & 40\\ 
\hline
\end{tabular}
\\
\end{table*}


\subsection{Infall region and backsplash galaxies}
\label{subsection:infall}

Fig.~\ref{sr-fig12} (top panel) shows the two clusters and their surroundings. The colour difference in the two environments is striking (mostly blue for Abell 1367 and mostly red for Coma). The two clusters  being connected to the same filament, 30 Mpc across, have diverse characteristics, not only in their dynamics but also the large-scale environment that surrounds them.

For each of the groups and the two clusters, we compute a statistical estimate of their virial radius, {(R$_{\rm virial}$)} \citep[e.g.][]{porter2005}, as 
\begin{equation} 
~\label{vir1}
r_v\sim\frac{\pi}{2}\frac{N(N-1)}{\sum_i\sum_{i<j}R_{ij}^{-1}},
\end{equation}

The Bottom panel shows the 1D distribution of group and non-group galaxies as a function of the virial radius from near the clusters (excluding the clusters). It shows that while both galaxies belonging to groups and non-group ones near the Coma cluster experience a steep decline in the star-formation from near its virial radius, the trend is almost flat near Abell~1367 for both the populations. The two dashed lines represent the range in cluster-centric radius
$1 < R/R_{\rm virial} < 2$, which is expected to have a mixture of galaxies infalling for the first time, and backsplash galaxies which are on their way back after a passage. 

There have been several estimates of the fraction of backsplash galaxies at various locations in a cluster. For instance, \citet{Pimbblet2011} estimated the backsplash fraction in the above range of radial distance to be around 50\%. Based on simulations, and spectral indices from the SDSS, \citet{Mahajan2011b} found about 20\% of these backsplash galaxies show signatures of recent or ongoing star formation. Given the limited scope of our analysis, if we were to remove 20\% of the star-forming galaxies in the radial range $1 < R/R_{\rm virial} < 2$ (accounting for the backsplash population) for the two clusters, which would lie in the environment bin [d], we would almost always get an over-estimation of the star-forming fraction of galaxies in this bin. We can thus conclude that, in this region, either the fraction of backsplash galaxies is less than 50\% or the contribution to the star-forming fraction from the backsplash population is greater than 20\%.

Since the star-forming fraction is even higher in non-groups, we can conclude that either the backsplash contamination is higher in this category, or a substantial part of the backsplash galaxies are star-forming. \citet{Mahajan2011b} has estimated the fraction of backsplash galaxies to be one-third at the projected virial radii of the cluster, and found the galaxies to be almost always quenched in just a single passage through the cluster. Our results are consistent with a lower fraction (less than 50\%) of backsplash contamination.

\section{Discussion}

In this paper, we investigate the dependence of the evolutionary properties (star-forming activity, nuclear activity and morphology) of galaxies, in a section of the cosmic web (the Coma Supercluster), on the local and very local environments of the galaxies. The smoothed projected density of galaxies surrounding each galaxy counts as the local environment, which is quantified using a kernel density estimator (KDE) algorithm. After smoothing, we divide the region into five distinct environment zones marked [a]--[e]. The properties of the groups to which the galaxies belong are considered for the very local environment. Our sample is split into two classes; the ones occupying a group halo (group galaxies) and the other outside such haloes (non-group galaxies). Group galaxies are further classified into those belonging to high-mass and low-mass group haloes, as defined by the parent catalogue of groups.

\somaked{The Coma Supercluster has two clusters, Coma (Abell~1656) and Abell~1367, diverse in their characteristics. The Coma cluster seems to be a dynamically relaxed cluster, mostly consisting of quenched red-sequence galaxies, Abell~1367 largely consists of blue and young late-type galaxies, contains multiple sub-groups \citep{Cortese2004}, and appears to be in the process of being assembled \citep{Moss1998}. }

\subsection{Evolution of galaxies: internal vs external mechanisms}

If galaxy evolution were passive, we would expect the internally driven mechanisms, for instance AGN feedback, to play a major role in the quenching of star formation, irrespective of environment. 
Comparing Fig.~\ref{sr-fig3} (star-forming fraction) and Fig.~\ref{sr-fig5} (morphological fraction) with Fig.~\ref{sr-fig6} (optical AGN fraction) and Fig.~\ref{sr-fig7} (radio AGN fraction),
 we find a stark contrast in the effect of group environment on the member galaxies. These figures indicate that while
 the group galaxies are preferentially quenched, and have early-type morphology,  compared to non-group galaxies,
 the group environment seem to favour optical and radio AGN activity. 
 The physical processes occurring in the group environment seem to favour nuclear activity over star formation, although both of them rely on the supply of cold gas for their sustenance. This indicates that the extra quenching of star formation seen in groups is partially due to AGN feedback in the groups. 
 
 Among the group galaxies, Figs.~\ref{sr-fig8} and \ref{sr-fig11} show that, on further splitting  into high and low richness groups, groups belonging to the high-richness groups have a lower AGN fraction that in low-richness groups.
  This might be due to the lower availability of the supply of cold gas in higher-richness groups \citep[e.g.][]{Hess2013} or the shock-heating in haloes of masses of 10$^{12}$ \Msun{} or higher \citep[e.g.][]{Birnboim2003, Dekel2006}. Thus, in high richness groups, the suppression of star formation is more linked to these effects than that of AGN feedback. On the other hand, in low-richness (low velocity dispersion) systems, galaxies experience enhanced tidal interaction over longer time-scales. Such processes predominantly fuel active nuclei but not necessarily enhance star formation at the same time \cite[e.g.][]{Gordon2018a}.

Turning to the galaxies which are not members of groups, Fig.~\ref{sr-fig6} shows that  in the densest local environment bin [e] 14\% of the non-group galaxies have optical AGN, compared to 10\% those in groups in [d] (containing the infall region of the Coma cluster, and the A1367 cluster, as well as the two rich groups), and in the Coma cluster.

The infall regions of the cluster tend to favour such activities due to tidal compression leading to collapse  of gas clouds \cite[e.g.][]{schulz2001}. We conclude that the local environment density is the primary driver of the relative abundance of nuclear activity, which can be further enhanced in an optimal form environment, where the gas content is sufficient and triggering due to tidal interaction can occur.

\subsection{How important is the group environment for galaxies infalling into clusters?}

In Fig.~\ref{sr-fig3} and Fig.~\ref{sr-fig4}, we noted a higher fraction of star-forming and late-type galaxies in groups in the environmental bin [d] than in [c], which deviates from the general decline of the SF fraction with KDE-smoothed density from [a]--[e] seen in non-group galaxies. In Fig.~\ref{sr-fig5}, we see that, despite the segregation in magnitude, both the bright and faint galaxies continue to follow the trend in terms of their star formation in all five environments. 
The expected continuous decline in group galaxies from environment bin [a] to [c] indicate a higher degree of pre-processing (enhanced quenching), which the group galaxies go through,  as the group passes through  progressively higher densities. In this section, we examine the reasons behind this reversal, and why this occurs only for the group galaxies. For this, we look at both mechanisms internal and external to groups.

Groups can  passively evolve due to physical processes that solely affect their constituents. There lower velocity dispersion compared to clusters results in higher dynamical friction between its galaxies, leading to enhanced tidal interaction and stripping of cold gas. Much of this gas is retained by the gravitational potential of the dark matter halo of the galaxy group, and it can be used to rejuvenate the star formation in some of its galaxies, which might have ceased earlier due to quenching processes. Thus in a majority of groups, a higher fraction of star-forming galaxies and late-type galaxies are favoured. \citet{Roberts2017} found that galaxies closer to the cores of non-relaxed groups are  more star-forming than in their relaxed counterparts. 

Turning to external factors, there is a possibility that the cold gas in groups is continuously replenished by accretion from the filaments in which they are embedded, in cold mode \citep[e.g.][]{Keres2009,Kleiner2017}, getting converted into stars. \citet{Vijayaraghavan2013} have investigated the possible effects of group-cluster mergers on group haloes, and on the galactic content of groups and clusters, using hydrodynamical simulations. They reported tidally distorted groups in the proximity of massive clusters, as far as 4~Mpc away from the cluster centres, the distortions increasing as the group falls further along the filament. However, with a closer look, they ruled out the possibility of tidal distortion taking place within the group's virial radius, and concluded that this does not result in  pre-processing. 
The confluence of supercluster filaments at the outskirts of clusters could also give rise to an environment which is conducive to rapid transformation of galaxies, involving star-forming activity, AGN activity and morphological transformation. 

Several supporting examples confirm these abrupt transformations in galaxy properties in  the infall zones of clusters, particularly those explicitly fed by filaments. For example, \citet{Gordon2018a} finds  substantial AGN populations at the periphery of  groups, as opposed to in their core regions. \citet{Gavazzi2011} observed the same phenomenon in and around the the Coma cluster. It has been suggested that in the filaments that feed clusters, rapidly infalling galaxies are confined to  narrow solid angles, resulting in enhanced galaxy-galaxy interactions,  leading to harassment-induced starbursts \citep[e.g.][]{porter2007,Mahajan2012}. This can also cause compression in the intragalactic gas, fuelling nuclear activity. 

\subsubsection{Group galaxies in the infall region of clusters}
\label{subsection:Impact of galaxies in groups in the infall region}
For galaxies belonging to groups travelling from the filamentary web of the Supercluster to the inner regions of a cluster, both the effects internal to the group, and those due to the external environemnt within a filament, are at play. We now examine which of these are more important in producing the enhancement seen in [d], over and above the natural decline in SF seen in galaxies from the filament environment to the core of the cluster?
By studying the galaxies on filaments falling into clusters, \citet{porter2008} found an enhancement in star-formation between R$_{\rm virial}$ and 2R$_{\rm virial}$ away from the cores of clusters, in a sample of 50 clusters fed by long inter-cluster filaments, an effect not seen in a comparable ``control" sample not connected to obvious filaments. This effect has been confirmed in subsequent studies, e.g., \citet{Ramatsoku2019,Liao2019,Vulcani2018}.
On further splitting this sample between non-group and group galaxies, the effect was found at the same distance from the cluster core, albeit more enhanced in group galaxies.  This indicates that the dark matter haloes of groups may help their galaxies retain their reservoir of cold gas necessary for enhanced star formation due to the effect of the filamentary environment in the infall region.

Comparing with this result from a sample of superclusters, in the Coma Supercluster that is being examined in this paper, we see opposing results at the two ends of the central filament
(Fig.~\ref{sr-fig12}). In the infall region of the Coma cluster (left panel of figure), we find a steady incline in the SF fraction as we move radially away from the cluster centre, till about 2R$_{\rm }$, whereafter it declines. On the other hand (right panel of figure) for the case of Abell 1367, we find a constant slope for the star formation, albeit a bit higher at 70\%, than Coma. Even though on an absolute scale we see significant star formation near Abell 1367, but on a relative scale the modulation in the star-forming activity due to the cluster environment, especially an enhancement just before the infall to the cluster, is pronounced in the case of Coma cluster. The trend further suggests that the galaxies on the outskirts of Abell~1367 are relatively more immune to the presence of the cluster. We discuss the enhancement seen in groups near Abell 1367 in greater detail in the next sub-section.

We see a similar effect of the presence of the Coma cluster on the non-group galaxies as well. In Fig.~\ref{sr-fig13}, in the [d] environment bin, we split the group and non-group galaxies near the two clusters to see whether the local overdensity, or the fact that the galaxy in in the infall region of the cluster, is more dominant. The [d] environment has about 25\% of its galaxies from the outskirts of Coma, the  and rest from the outer regions of Abell~1367. We observe that despite being in the same density bin [d], the non-group galaxies near Coma are much more enhanced in star formation than those near Abell~1367. Note that the [d] environment in case of Coma forms the next encircling contour while in the case of Abell 1367, it forms the end contour and contains Abell 1367 and two other prominent groups (see Fig.~\ref{sr-fig2}). This again implies that for galaxies outside the group environment, the position relative to the cluster matters more than their local overdensity. 

We should discuss a caveat here. We may expect a systematic contamination, particularly in the crowded [d] environment near Abell 1367, due to projection effects. Our designation of group and non-group galaxies comes from the Yang et al. catalogue, whereas the definition of environments comes from our KDE analysis of a two-dimensional catalogue in RA and Dec (albeit in the velocity range 4500-9500 km/s). This region has two massive groups and the cluster Abell~1367. We have looked closely at the ``non-group" galaxies (according to the Yang et al. criterion), and find that the contamination to the overall sample of ``non-group" galaxies, from galaxies whose velocities are outside the $\pm 3\sigma$ limits from the mean of each group/cluster, is not significant.

In order to estimate the likely contamination from the background or foreground, the percentage of  ``non-group" galaxies which are outside $\pm 3\sigma$ of the mean radial velocity of each group (within their ``virial radius" as defined in \S{\ref{subsection:infall}}), is  4\% and 8\% of all ``non-group" galaxies, in the richer [d] and [e] environment bins respectively. If we take all galaxies only within the volume defined by the virial radii of all designated groups in the  [d] and [e] environments, even then this fraction is 34\% and 31\% respectively.

In the HI detected galaxies from the (ALFALFA survey; \cite{Haynes2018}), in the same region of the sky as ours, \cite{Hess2013} show that the HI mass to the stellar light ratio (HI/L$_{r}$) of galaxies within 5~Mpc of the Coma Cluster, 
is substantially reduced  compared to  elsewhere in the Supercluster. Within two virial radii of the Coma cluster, the HI detected field galaxies were found to retain significant cold gas content, compared to their group counterparts. This is consistent with the effect seen in the [d] bin in this work. 

\subsection{Excessive star-formation in groups near Abell 1367}
\label{excess_SF}

In this section we take a closer look at the differences between the two ends of the Coma Supercluster, in particular the infall regions of the two clusters clusters that contribute to the [d] environment bin.

 In the region around Abell 1367, which is part of the environment bin [d], it turns out that a major contribution to the fraction of  star-forming galaxies comes from the two massive groups located in the outskirts of Abell~1367 (see top right-hand panel of  Fig.~\ref{sr-fig12}), which together contribute 69\% of star-forming galaxies. In contrast, in the infall region of the Coma cluster at the other end of the filament, the groups only have 31\% star-forming galaxies.
In general, in the environment bin [d], the star-forming fraction in groups is about 58\%, compared to fractions of 48\% and 62\% in [c] and [b] respectively.  Clearly the groups in the vicinity of Abell 1367 have a higher-than-average rate of star-formation, almost equivalent to that in the field population.

Referring to Figs.~\ref{sr-fig1}, \ref{sr-fig2} and \ref{sr-fig12}, we find that these two rich groups are the NGC~4065 group, known to be a dynamically un-relaxed group \citet{Helsdon2000,Forbes2006}, and the NGC~3937 group, with galaxy richness of 160 and 70 respectively in our group catalogue. They lie at distances of 7.6 Mpc (NGC 4065) and 3.5 Mpc (NGC 3937) from Abell~1367. \cite{Gregory1978} were the first to point out the presence of massive groups near Abell 1367, and the complete absence of such systems in close proximity of Coma. As our observations indicate excessive star formation in all these systems, we can conclude that they are dynamically young systems  \citep[e.g.][]{Bekki1999, Poggianti2004, Cortese2004, Roberts2017}. A third group, also with enhanced star formation, previously noted by \cite{Cortese2006} to be falling into the cluster, famously called BIG (blue infalling group, \cite{Gavazzi2003A}) also lies in this region, though it is not included in our analysis due to our magnitude cut. It has been suggested in many studies that Abell 1367 is a dynamically unrelaxed cluster \citep[e.g.][]{Moss1998},
which, together with nearby groups, forms a system which is collapsing for the first time and is much younger than the  Coma cluster system at the other end of the Supercluster. We look for clues for such a stark difference between the two systems.

The fraction of star-forming galaxies in  Abell~1367 is 50\%, compared to 34\% in the Coma cluster. This is consistent with the observation \citep[e.g.][]{Mahajan2012}  that dynamically unrelaxed clusters have a higher mean SFR than relaxed ones, at all values of cluster-centric radius, and their member galaxies tend to have a higher mean local density in their surroundings. 

If star formation relies on infalling ``cold mode" gas from the larger system of filaments of the cosmic web surrounding the Coma Supercluster, in groups and clusters this accreted gas can get shock-heated (in systems with  halo mass $>10^{12}\,M_\odot$) to nearly virial temperatures (millions of K), due to the deep potential offered by the dark matter halo  \citep{Birnboim2003}. Thus in these systems the gas becomes unavailable to form stars  \citep[e.g.][]{Keres2009}. Since both the groups in the vicinity of Abell~1367 are massive ( 14.2\M{} and 13.9 \M{}), and are high in the content of  star-forming galaxies, as well as rich in H{\sc i} \cite{Freeland2009}, there is a likelihood that the acquired gas in these groups has cooled down.

The growth of clusters is highly dependent on the surrounding cosmic web and availability of matter present to feed these massive structures.
As we have seen in our analysis (see \S\ref{section:Section-3}), the overall shape of the underlying galaxy density apparent in Fig.~\ref{sr-fig2}, in particular the contour confining the [d] region, covers a wider area near Abell 1367 than Coma. While Abell 1367 is still collapsing, Coma seems to have already engulfed infalling substructures from its immediate vicinity \citep[e.g.][]{Burns1994}.

From the work of \citet{Weinmann2006,Woo2013,Paranjape2018}, for example, it is apparent that the mass of the group-scale halo is the prime driver of its properties, and is sufficient to account for the clustering properties of galaxies, and other properties of the constituent galaxies such as colour and sSFR. This leads to  `one-halo conformity' \citep{Weinmann2006}. Another similar hypothesis, which is based on conformity, but is not limited to one halo, extending much further than the defined virial boundary of the halo (`two-halo conformity') \citep{Kauffmann2013}, is due to a process known as `assembly bias' \citep[e.g.][]{Miyatake2016,Musso2018}. 

Fig.~\ref{sr-fig12} seems to favour connections beyond a single-halo hypothesis as the population in groups seems to have been derived from the same underlying colour distribution suggestive of a possible connection across haloes.
The colour distribution seen in galaxy population is not statistically favoured by their massive group haloes. This implies that these groups are statistically different from the expected population of groups in this mass range.

\section{Conclusions}
\label{section:Section-6}

In this paper, we have looked closely at effect of the group environment, compared to other measures of environment, in the evolutionary history of galaxies. Our subject was the Coma Supercluster, which is a system of clusters, groups and filaments about 100~Mpc away. We used spectroscopic parameters for galaxies from SDSS DR13 and a galaxy group catalogue derived from the SDSS \citep{Yang2007}. The sample was first divided into five classes of environment, using a kernel density estimator (KDE), based on the population over-density across the supercluster. Further, the sample was divided into group/non-group galaxies, to study the role of environment in determining the evolution of galaxies as a function of halo properties and large-scale environment. 
\\
The key results of our analysis are as follows:
\begin{enumerate}
\item Galaxies undergo pre-processing in groups in the Local Universe, independent of the location and halo mass of the groups. The degree of pre-processing can vary with its location in the large-scale environment (Figs.~\ref{sr-fig3} and \ref{sr-fig5}).
\item  We found evidence of evolutionary trends, including enhanced suppression of star-formation, and the lack of late-type morphology, with increase in galaxy density. 
\item We observed that over and above a general trend of decline in the star-forming fraction, with increasing over-density, the groups showed an enhanced presence of SF galaxies in the most over-dense environmental bin, right in the infall regions just outside the virial radii of clusters. This trend however was not present in the non-group galaxies.
\item Our results are consistent with cold mode accretion of gas into the groups from the filaments of the cosmic web being responsible for enhancing star formation in galaxies in the group environment. It appears that the group environment can hold such gas for longer and make it available to its galaxies.
\item These systematic trends were present for both massive (bright, $M_{r} < -18.5$) and low mass (faint, $M_{r} > -18.5$) galaxies (Fig.~\ref{sr-fig4}).
\item We found that at the current epoch, mostly the low mass galaxies are the ones forming stars, irrespective of their local densities as well as group/ non-group membership \citep[e.g.][]{Haines2007}.
\item We find a higher fraction of AGNs at both optical and radio wavelengths to be present in galaxies in groups as compared to the isolated galaxies (Figs.~\ref{sr-fig6} and ~\ref{sr-fig7}).
\item we find that the low-richness groups have a higher star-forming, late-type morphology and AGN fraction than their high-richness counterparts (Figs.~\ref{sr-fig8}, ~\ref{sr-fig10} and ~\ref{sr-fig11}).
\item{Low-richness groups tend to favour (optical) AGN activity in their galaxies, compared to galaxies in high-richness groups, or non-group galaxies}. AGN activity is enhanced also in non-group galaxies in the infall regions of clusters, where the filament overdensity favours the triggering of AGN.
\item We report the exceptional presence of at least two high richness groups, lying on the outskirts of Abell 1367, with a high fraction of star-forming and late-type galaxies, present in the environmental bin with highest density (Figs.~\ref{sr-fig8} and \ref{sr-fig10}). This trend was not seen for the low-richness groups. This is inconsistent with \citep{Wetzel2012}, who had shown that an increase in group richness would lead to quenching.
\item The surroundings of the two clusters, Coma and Abell 1367, are distinctive in their galaxy content. The galaxies present in the vicinity of Abell 1367 are mostly star-forming (blue), while in case of Coma they were mostly quenched (red)  (Figs.~\ref{sr-fig12} and \ref{sr-fig13}).
\item We believe that Abell~1367 is in the process of assembly, in a distinctive zone of the cosmic web, where group/cluster formation is still ongoing. We suspect all these groups are not completely virialised and are still accumulating its constituents from the surrounding cosmic web, implying that the assembly is highly correlated with the underlying large-scale environment, especially in the cluster infall regions.
\item High-mass group haloes and high local galaxy density can lead to unusually high star-formation rates in exceptional circumstances (as in the groups on the outskirts of Abell 1367).
\item In this Supercluster, We find substantial agreement with a two-halo theory, where the similarities in galaxy properties within a halo, were also common with those in its neighbouring haloes (Fig.~\ref{sr-fig12}). 
 
\end{enumerate}

\section*{Acknowledgements}
We thank the anonymous referee for a thorough review of our manuscript and constructive comments which have significantly helped improve this paper. RS thanks the University Grants Commission, India, for financial support. We thank Omkar Bait, Nikhil Mukund, Raja Guhathakurta, Joel Primack, Aseem Paranjape, Gulab Dewangan, Joydeep Bagchi and Smriti Mahajan for their useful suggestions and comments. This research work used NASA's NED and ADS services, and made use of Numpy, Astropy and Scipy packages in Python, and the TOPCAT software developed by the University of Bristol and the erstwhile STARLINK. SDSS is run by the Sloan Foundation and its participating Institutions, NASA, NSF, the Japanese Monbukagakusho and the Max Planck society. The SDSS website is \url{https://www.sdss.org/}.

\section*{Data Availability}
The data underlying this article are available in the article and in its online supplementary material.
%
\bibliographystyle{mnras}
\bibliography{references.bib} 

\bsp    

\section*{Appendix A1: Full Table of galaxy groups present in the Coma Supercluster}
 
Available with the online version of this paper.
\label{lastpage}
\end{document}